\documentclass[usenatbib]{mn2e}
\usepackage{color}
\usepackage{natbib}
\usepackage{paralist}
\usepackage{graphicx}
\usepackage{epstopdf}
\usepackage{epsfig}
\usepackage{amsmath}
\usepackage{tablefootnote}

\def\ciso{{$^{12}$C/$^{13}$C}}

\def\teff{\mbox{$T_{\rm eff}$}}
\def\logg{\mbox{log~{\it g}}}
\def\vmicro{\mbox{$\xi_{\rm t}$}}
\def\kmsec{\mbox{km~s$^{\rm -1}$}}

\def\eg{\mbox{e.g.}}

\title[Chemical compositions of RGs in the 
NGC 752]{THE CHEMICAL COMPOSITIONS AND EVOLUTIONARY STATUS OF RED GIANTS 
IN THE OPEN CLUSTER NGC 752}
\author[G. B\"{o}cek Topcu, M. Af\c{s}ar, M. Schaeuble and C. Sneden]{G. 
B\"{o}cek Topcu$^{1,2}$\thanks{E-mail: gamzebocek@gmail.com (GBT); 
melike.afsar@ege. edu.tr (MA); mschaeu@astro.as.utexas.edu (MS); chris@verdi.as.utexas.edu (CS)}, 
M. Af\c{s}ar$^{1,2}$, M. Schaeuble$^{2}$, C. Sneden$^{2,1}$\\ 
$^{1}$Department of Astronomy and Space Sciences, 
                 Ege University, 35100 Bornova, \.{I}zmir, Turkey;\\
$^{2}$Department of Astronomy and McDonald Observatory,
                 The University of Texas, Austin, TX 78712}
\begin{document}

\date{Accepted 2014 November 7 }

%  \pagerange{\pageref{firstpage}--\pageref{lastpage}} \pubyear{XXX}

\maketitle

\label{firstpage}

\begin{abstract}
We present detailed chemical compositions of 10 red giant
star members of the Galactic (open) cluster NGC~752, derived from
high-resolution (R $\approx$ 60,000), high signal-to-noise 
($S/N$ $\geq$ 140) spectra.
We confirmed cluster memberships by measuring the stellar radial 
velocities, and by deriving model atmosphere parameters 
(\teff, \logg, [Fe/H] and $\xi_{t}$) from equivalent widths of
\mbox{Fe\,{\sc i}}, \mbox{Fe\,{\sc ii}}, \mbox{Ti\,{\sc i}}, and 
\mbox{Ti\,{\sc ii}} lines.
The metallicity we obtained for NGC~752 ([Fe/H]~=~$-$0.02$\pm$0.05) is in good agreement with previous
studies. 
We derived abundances of $\alpha$ (Mg, Si, Ca), light odd-Z (Na, Al), 
Fe-group (Sc, Ti, V, Cr, Mn, Fe, Co, Ni, Cu, Zn),
$n$-capture (Y, La, Nd, Eu), and $p$-capture (Li, C, N, O) species
for each star. 
Furthermore, we also derived abundances
of the LiCNO $p$-capture element group and carbon isotopic ratios,
using synthetic spectrum analyses of the \mbox{Li\,{\sc i}} 6707~\AA\ 
resonance doublet, the [\mbox{O\,{\sc i}}] line at 6300~\AA, the CH G-band 
features near 4311 and 4325 \AA, the C$_{2}$ bandheads at 5160 and 5631 \AA,
and $^{12,13}$CN red system lines in the 7995$-$8040~\AA\ region.
By applying recent isochrones to NGC~752 photometry, and comparing
the color-magnitude diagram information to our Li abundances and \ciso\ 
ratios, we suggest that the 10 observed red giants can be separated into
three first-ascent, six red-clump and one red horizontal branch star. 

\end{abstract}


\section{Introduction}\label{intro}

Open clusters (OCs) are excellent probes to investigate both 
stellar and Galactic disk evolution. 
According to classical theories, OCs members were formed from the 
same protocluster cloud at the same time and the same distance. 
This lends particular importance to the morphology of an OC 
color-magnitude diagram (CMD), as it allows determination of the 
temperatures, luminosities, and evolutionary states of cluster members 
more accurately than can be done for field stars.

All members of individual OCs should have comparable initial chemical 
compositions, differing only in their initial masses. 
As predicted by the models, the individual nucleosynthesis and mixing
histories of the members are responsible for any surface abundance 
differences among evolved OC stars. 
In classical stellar evolution, the so-called first dredge-up, which occurs 
at the base of the red giant branch (RGB)\footnote{
In this paper we will use the term red giant (RG) generically, meaning
all cool and luminous stars; RGB will designate stars on the first ascent 
of the giant branch; and RC will designate He-burning red clump stars.}, 
is the main mechanism that changes the surface abundances.
The best indicators of this mixing are the observed changes in abundances
of the elements susceptible to proton-capture mechanisms.
In stars with masses and metallicities similar to the sun, photospheric 
abundances of $^{12}$C and \ciso\ will decrease while 
the abundances of  He (unobservable), $^{13}$C and $^{14}$N will increase. 
Surface Li abundances vary greatly from one star to the next, and on
average will also greatly decrease during the star's evolution from main sequence to the RGB phase.  

The quantitative changes in LiCNO abundances depend on the initial masses 
and metallicities of the stars (e.g. \citealt{swe89,char94,boo99,mar01}).
For example, standard predictions suggest that \ciso~= 20 to 30 is 
typical for  RGB stars (e.g. \citealt{iben67,dear76}),  
though observations yield a more complex answer. 
In particular, early spectroscopic studies 
(e.g. \citealt{tom75,tom76,lambert81,sne86b}) showed that RGB and RC 
stars often exhibit \ciso~$<$~15, much lower than expected.

\citeauthor{gil89} (1989, hereafter Gil89) found a strong 
anti-correlation between stellar mass and \ciso\ ratios in clusters 
with turnoff masses M $<$ 2.2 M$_{\odot}$.  
This trend could not be explained by canonical stellar evolution models 
without invoking some sort of extra convective envelope mixing. 
Recently, \cite{afs12} have discovered anomalously low \ciso\ ratios among 
metal-rich thin-disk field red horizontal branch stars, indicating that their 
carbon isotope ratios may have been altered during RGB evolution. 
Several extra-mixing processes have been suggested to explain these low 
isotopic ratios, \eg, \cite{swe79,char94,char98,boo99,char10}. 
For first-ascent RGB stars with M $<$ 2.2 M$_{\odot}$, especially 
at low metallicities, both \ciso\ and C/N ratios sharply drop at the so-called 
``luminosity function bump" \citep{grt00}. At this evolutionary stage, the 
outward-advancing H-burning shell cancels out the chemical discontinuity left by 
the convective envelope, allowing further mixing to take place (e.g. 
\citealt{char94}).

The observational challenge is to discern the results of the extra-mixing 
processes through the abundances in stars with well-determined masses 
and luminosities. This is difficult for field stars, since it is not easy to determine 
their masses. However, mass estimation is easily done for open and globular 
cluster RGs. Detailed studies of LiCNO abundances of evolved members of 
individual OC's are increasing (\eg,
\citealt{gil89,gil91,luck94,taut00,taut05,smil09,miko10,miko11a,miko11b,miko12}).
In this paper we aim to add to this growing literature by presenting 
high-resolution spectral analyses of NGC~752 RG member stars. 
We report atmospheric parameters, [Fe/H]
metallicities\footnote{
For elements A and B,
[A/B] = log $(N_{A}/N_{B})_{\star}$ -- log $(N_{A}/N_{B})_{\sun}$
and log $\epsilon$(A) = log $(N_{A}/N_{H})$ + 12.0 .
Also, metallicity will be taken to be the [Fe/H] value.}, 
relative abundances and ratios of elements belonging to the
$\alpha$,  light odd-Z, Fe-peak, and neutron-capture groups, 
with a particular focus on the LiCNO group.
We employ new, comprehensive laboratory transition studies 
of CH, C$_2$, and CN molecular bands, which have materially increased the 
reliability of proton-capture abundances derived in RG stars.

The structure of this paper is as follows: 
in \S\ref{cluster752} we introduce NGC~752 and its RGB members.
The observations and reductions are outlined in \S\ref{obsredEW}.
We discuss compilation of atomic/molecular line lists and equivalent width 
measurements in \S\ref{lineEW}.
The derivation of model atmospheric parameters is described in \S\ref{models},
followed by abundance analysis in \S\ref{abunds}.
Finally, in \S\ref{discuss} we discuss the implications of our results
for the evolution of NGC~752.

\section{NGC~752 and Its RG Stars}\label{cluster752}

NGC~752 is one of the closest ($\simeq$447~pc, \citeauthor{dan94} 1994, hereafter Dan94) intermediate age 
(1$-$2 Gy) OCs.
Due to its age and distance, NGC~752 has been the focus of
many photometric and spectroscopic 
studies for nearly a century.
Basic parameters of this cluster are given in Table~1. We will
elaborate on some of these quantities below.

\begin{table}
 \centering
 \begin{minipage}{95mm}
 \label{tab1}
  \caption{NGC 752 cluster parameters.}
  \begin{tabular}{@{}llc@{}}
  \hline
   Quantity    &
   Value     &
   Ref.  \\ 
 \hline
Right Ascension (2000)   &       01 57 41   & WEBDA \\
Declination (2000)       &    $+$37 47 06 & WEBDA \\
Galactic longitude       &        137.125 & WEBDA \\
Galactic latitude        &      $-$23.254 & WEBDA \\
Distance                 & 447 $\rm {pc}$ & Dan94\\
$E(B-V)$                 &          0.035& Dan94 \\
$(m-M)_0$                  &           8.25& Dan94 \\
Age                  &          1.7 to 1.9 Gyr  & Dan94\\
Turnoff Mass        &          1.5 M$_{\odot}$ & \cite{bart07} \\
\hline
\end{tabular}
\end{minipage}
\end{table}

Photographic observations and member identifications  
of NGC~752 were first made by \cite{hein26}.
\cite{ebb39} conducted the initial proper motion study of
this cluster and also expanded the membership identifications
of \cite{hein26}.
Using new stellar identification systems, the membership of several 
other stars in NGC~752 has been established.
\cite{mer98} summarized past numbering systems for the 
RG members of NGC~752 and confirmed the membership of 15 of these stars.
Four of them have been reported as spectroscopic binaries and were thus not 
included in our sample.
Later, \citeauthor{mer08} (2008, hereafter Mer08), using the  \cite{hein26} numbering 
system, introduced a new naming system that we have adopted 
through this study, using the prefix ``MMU''.

\cite{mer98} concentrated on the binary frequency and the
RG branch morphology of NGC~752.
Investigating the color-magnitude diagram (CMD) of the cluster, they 
discovered an unusual distribution of RGB stars. 
They showed that although the RGs nominally occupy
the so-called He-burning clump domain, some of them were 
fainter and bluer than true clump members ought to be.
\cite{gir00} investigated this unusual morphology in 
NGC~752 and NGC~7789, which has a similar red clump distribution. 
They suggested that the peculiar distribution can occur if OCs 
have He-burning stars with a significant mass spread at the same time. 
The lower-mass stars would have undergone standard He-flash prior to
settling in the clump, while the higher-mass stars would have ignited helium 
quiescently.
They also suggested that the dispersion in red clump masses might have been
caused by different amounts of mass-loss that had taken place during the 
RGB phase.

Detailed abundance analysis of individual RC members in this cluster began 
with \cite{plw88}, who  investigated the lithium abundances of 
evolved stars in NGC~752 and M~67 using the \mbox{Li\,{\sc i}} 6707 \AA\ 
resonance doublet.
They observed 24 RGs in the vicinity of NGC~752 and reported 
Li abundances for 11 members previously identified by \cite{hein26}.
Among those, only Heinemann 77 and 208 yielded strong Li features with 
abundances of log$\epsilon$(Li)~=~1.1 and 1.4, respectively.  
These stars were thus interpreted as first-ascent giants, while the rest of 
the observed members were identified as evolved He-burning clump stars. 
As mentioned in \S\ref{intro}, Gil89 reported high-resolution spectral 
analyses of C isotopic ratios from CN features near 8004~\AA\
and new Li abundances in selected RGB members of 19 clusters. 
In the lower turnoff mass OCs of her sample, a clear correlation
between cluster turnoff mass and \ciso was found.
The prior results for NGC~7789 (\citealt{pil86}, \citealt{sne86a}) fit in with
the general trend; see Gil89 Figure~9.

Interpretations of NGC~752 RG evolutionary states depend 
on fundamental cluster parameters such as distance
modulus, reddening, age, main-sequence turnoff mass, and metallicity.
Here we summarize the literature values of these quantities, noting
that estimates of $(m-M)_{0}$, $E(B-V)$, age, $M_{TO}$, and
[Fe/H] have not significantly changed from earliest investigations 
to the present time. 

\textit{Reddening and Distance Modulus:}
In the publication history of NGC~752, most of the methods for deriving 
these quantities have been applied to its main sequence members. 
Dan94 collected the data of six photometric 
systems and transformed them to the Johnson UBV system. 
They derived $E(B-V)$~= 0.035$\pm0.005$ and $(m-M)_{0}$~= $8.25\pm0.10$ 
(447$\pm$10~pc).
\cite{bart07} conducted seven-color Vilnius photometry for NGC~752, and from  
the best isochrone match to the CMD they found $(m-M)_{0}$~= $8.06\pm0.03$ 
(409$\pm$6~pc).
Later, \cite{bart11} analyzed Vilnius multicolor photometry of NGC~752 
with an expanded sample that extended down the main sequence to $V$~$\cong$~18.5.
Using the data of 70 photometric members, they derived 
$(m-M)_0$~= $8.37\pm0.32$ (472$\pm$72~pc), consistent with their prior
estimate due to its larger uncertainty.

\textit{Cluster Age and Main Sequence Turnoff Mass:}
These quantities have been determined by comparing the NGC~752 
photometry to isochrones.
Dan94 derived an age of  $1.9\pm0.1$ Gyr from 
matching classical isochrones, and $1.7\pm0.1$ Gyr from isochrones 
that included convective overshooting.
The isochrone matches to Vilnius-sytem photometry yielded an age of 
$1.58\pm0.04$~Gyr \citep{bart07} and $1.41$~Gyr \citep{bart11}.
 the cluster turnoff mass \cite{bart07} suggest 
M$_{TO}$~$\approx1.5$ M$_{\odot}$, while
Gil89 estimated a turn off mass M$_{TO}$~$\approx1.6\pm0.13$ M$_{\odot}$ 
using photometric data then available.

\textit{Metallicity:}
Since metallicity is one of the most important parameters in the evolution of 
OCs, several NGC~752 studies have been devoted to this parameter.
Like most OCs, its metallicity is approximately solar. 
Nevertheless, a variety of photometric and spectroscopic techniques have 
been applied to derive the [Fe/H] value more accurately.
Dan94 used both spectroscopic and photometric approaches to derive a mean
cluster metallicity of [Fe/H]~= $-0.15\pm0.05$.
From Vilnius photometry, \cite{bart07} derived 
$\langle$[Fe/H]$\rangle$~= $-0.12\pm0.03$, and \cite{bart11} 
revised this value to 
$\langle$[Fe/H]$\rangle$~= $+0.16\pm0.09$.
\cite{pp10} developed a statistical method to determine OC 
metallicities, finding $\langle$[Fe/H]$\rangle$~=~$+$0.05.
\cite{pau10} derived $\langle$[Fe/H]$\rangle$~= $-0.15\pm0.11$ by using the 
previously published photometric data sets of the cluster. 
\cite{two06} reported Str{\" o}mgren photometry for 7 RG and 
21 main-sequence stars of NGC~752, and from 10 non-binary members they 
derived [Fe/H]~=~$-$0.06~$\pm$~0.03.
Finally, \cite{hei14} reported a mean metallicity of 
$\langle$[Fe/H]$\rangle$~= $-0.02\pm0.04$, based on spectroscopic 
data gathered from 19 different sources.
There are two recent studies with detailed high-resolution spectral 
analyses of NGC~752. 
Based on four RGBs, \citeauthor{car11} (2011, hereafter Car11) determined a mean cluster metallicity of 
$\langle$[Fe/H]$\rangle$~= $0.08\pm0.04$. 
\citeauthor{reddy12} (2012, hereafter Red12) have reported a value 
$\langle$[Fe/H]$\rangle$~= $-0.02\pm0.05$
also from four RGBs, but only one is common with the sample of Car11. 
We conclude that a cluster metallicity of [Fe/H]~$\sim$~0.0 is a 
reasonable summary of past metallicity investigations.

For this study, we have adopted reddening, distance modulus, 
and age of NGC~752 from Dan94; see Table~1.
Fortunately, the above discussion suggests that the important cluster
quantities are generally agreed upon by all the literature 
studies of NGC~752.
For the NGC~752 turnoff mass (also
quoted in Table~1), we have adopted the value from \cite{bart07}, which
is one of the most recent, complete, and internally-consistent photometric studies of NGC~752.

\subsection{Colour--Magnitude diagram}\label{cmd}

\begin{figure}
  \leavevmode
      \epsfxsize=8cm
      \epsfbox{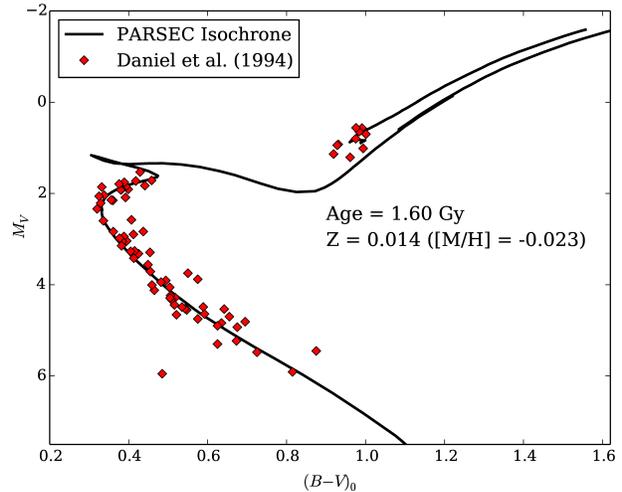}
       \caption[The Hubble Diagram]{The CMD of the OC NGC 752 with PARSEC isochrones \citep{bress12}.}
     \label{fig1}
\end{figure}

The CMD of NGC~752 has been studied by several authors 
(e.g. Dan94, \citealt{bart07} and \citealt{bart11}).
We used the combined photometric data
from Dan94, since the data from \cite{bart07} and \cite{bart11} do not include all the NGC~752 RG 
members present in our target list (Table~2).  

Figure~1 shows a comparison of the photometric data obtained by 
Dan94 to the latest set of PARSEC isochrones (\cite{bress12}). 
Earlier studies of NGC~752 employed isochrones published by \cite{van06}, 
\cite{mey93} and \cite{cas92} to obtain metallicity and age estimates 
of this cluster. 
However, several important physical considerations, such as convective 
overshooting, which might be important for age estimation in intermediate-age 
OCs such as NGC~752, have been neglected in these older calculations.
Additionally, old opacity data \citep{rod92,mea91} was used in the calculations 
of the final isochrone tracks.

\begin{table*}
\label{tab2}
 \centering
 \begin{minipage}{140mm}
  \caption{Observed NGC~752 RGs. Coordinates, K magnitudes and proper motions are taken from Simbad database. B and V magnitudes are from Dan94.} 
  \begin{tabular}{@{}lcccccccc@{}}
  \hline
   Star & Other Name  &   RA  & DEC  &  $B$ &  $V$  &  $K$  &  $\mu_\alpha$  &
   $\mu_\delta$       \\  & & (2000)   &  (2000)   & & & &   (mas/yr)  &  (mas/yr)     \\               
 \hline
\multicolumn{9}{c}{Cluster Members}    \\
MMU~1 	    &	BD+37 407  & 01 55 12.616 	&	$+$37 50 14.54	&	10.31	&	9.39	&	7.23	&	3.4    &  $-$10.8			  \\
MMU~3 	    &	BD+37 408  & 01 55 15.292	&	$+$37 50 31.29	&	10.41   &	9.47	&	7.20	&	11.3  &  $-$11.3		    \\
MMU~11   	& 	BD+37 409  & 01 55 27.669	&	$+$37 59 55.25	&	10.17	&	9.34	&	7.04	&	8.7    &  $-$13.2		   \\
MMU~24 	    & 	BD+37 412  & 01 55 39.370	&	$+$37 52 52.49	&	9.88	    &	8.90	&	6.55	&	7.5    &  $-$13.7   \\
MMU~27 	    &	BD+36 350 & 01 55 42.387	&	$+$37 37 54.55	&	10.13	&	9.17	&	6.80 &	6.9    &  $-$10.3	   	\\
MMU~77 	    &	BD+36 358  & 01 56 21.634	&	$+$37 36 08.37	&	10.54	&	9.43	&	6.92	&	4.8    &  $-$14.8	   	\\
MMU~137 	&	BD+37 424  & 01 57 03.110	&	$+$38 08 02.62	&	9.89 	&	8.89	&	6.64	&	8.7    &  $-$12.9		    \\
MMU~295 	&	BD+37 441  & 01 58 29,820	&	$+$37 51 37.60	&	10.23	&	9.29	&	7.04	&  9.4    &  $-$10.3	   	 \\
MMU~311 	&	BD+37 448  & 01 58 52.903	&	$+$37 48 57.23	&	10.07	&	9.04&	6.64	&	9.75   & $-$11.1			      \\
MMU~1367 	&	BD+37 450  & 01 59 14.800	&	$+$38 00 55.31	&	10.02	&	8.99	&	6.65	&	10.9   & $-$9.6			       \\
\multicolumn{9}{c}{Non Members}       \\
MMU~39 	    &	BD+37 415  & 01 55 54.759	&	$+$37 52 00.18	&	9.05 	&	8.09	&	5.77	&	$-$1.3  &  $-$11.3		    \\
MMU~215 	&	BD+36 370  & 01 57 42.736	&	$+$37 40 11.35	&	8.35 	&	7.12	&	4.45	&	15.0      &$-$32.5		  \\
 \hline
\end{tabular}
\end{minipage}
\end{table*}

PARSEC isochrones are calculated using not only two different types of 
overshooting (core convective \& envelope), but also two different types of 
opacities. For the high temperature regime, OPAL 1996 \citep{igl96} data is 
used in the computation, while the low temperature opacities are obtained 
from the AESOPUS code \citep{mar09}. This ensures that isochrone tracks for 
\textit{any} given input chemical composition can be calculated. In addition to 
these improvements, the equations of state were calculated from the lastest 
version of the freely available 
FREEEOS\footnote{http://freeeos.sourceforge.net/} package. The combination 
of these updated parameters and input physics should yield more reliable 
isochrones and thus also age and metallicity estimates.

To obtain our final estimates of the age and metallicity of this cluster, an 
iterative fit of the PARSEC isochrones to the photometric data was done. The 
final results can be seen in Figure~1. 
Our results are in good agreement with those obtained by previous studies, 
indicating that the choice of the isochrone source is not important.
The final input parameters of our isochrone are Z = 0.014 and an age of 1.6 Gyrs. 
These lead to a turnoff mass of M$_{TO}$~$\approx1.6M_{\odot}$, again in good 
agreement with all previous studies.

\subsection{RG target selection}\label{target}

To assemble the NGC~752 sample we began with the 
the OC database WEBDA\footnote{
http://www.univie.ac.at/webda/webda.html}, selecting stars 
labeled as RGs in its $V$ versus $B-V$ CMD.
We checked the membership status of these stars using the 
OC RG radial velocity (RV) survey of Mer08. 
In that paper, velocities were reported from the COROVEL ``spectrovelocimeter'' 
observations of 1309 potential members of 166 clusters.
Mer08 surveyed 30 possible members of NGC 752, finding just 10
stars without obvious spectroscopic binary companions and with RVs close 
to the cluster mean of $\langle RV\rangle$ =~$+$5.04~$\pm$~0.08 \kmsec\ 
(standard deviation of the sample $\sigma$~=~0.32~\kmsec).
We gathered spectra of these 10 stars.
We also observed two suspected non-member stars from their list to test
whether we could rule out their NGC~752 membership from our data alone.
Our target RGs are listed in Table~2, giving 
identifications in the MMU (Mer08) and BD systems.
Table~2 also lists RA, DEC, and proper motions values taken from 
the SIMBAD\footnote{
http://simbad.u-strasbg.fr/simbad/} 
database, B and V magnitudes from Dan94, and K magnitudes from
the 2MASS survey \citep{skr06}.
All of our programme stars are bright ($V$~$<$~9.47), thus easily accessible
to high-resolution spectroscopy.

\section{\textbf{Observations and Reductions}}\label{obsredEW}

High resolution spectra of the NGC~752 targets were gathered with 
the Robert G. Tull Cross-Dispersed Echelle spectrograph \citep{tull95} on 
the 2.7 m Harlan J. Smith Telescope at McDonald Observatory.
The wavelength range of the spectra was 4000 to 8000\AA\ with 
a resolving power of $R\equiv\lambda/ \Delta\lambda\approx60,000$. 
The spectral order coverage of this setup is continuous up to 
$\lambda$~$\simeq$~5700 \AA, but has small gaps between the echelle orders 
as the orders stretch in length at longer wavelengths. In Table~3 
we list the observation dates and exposure times for each of our target stars.
Three exposures per target were taken at each night. Exposure times 
given in Table~3 give the time for a single exposure.

\begin{table*}
\label{tab3}
 \centering
 \begin{minipage}{150mm}
  \caption{Observing logs and radial velocities of the observed stars.}
  \begin{tabular}{@{}llcccccc@{}}
  \hline
   Star  & Obs. Date  &  Exp.  &  S/N  &  $RV$\footnote{This study.}  & $RV$\footnote{\citet{mer08}.}  & $RV$\footnote{\citet{car11}.} & $RV$\footnote{\citet{reddy12}.}\\
 &  & (\textit{s})  & &($km s^{-1}$)  &($km s^{-1}$) &($km s^{-1}$) &($km s^{-1}$) \\                  
\hline
\multicolumn{6}{c}{Cluster Members}\\
MMU~1 	    &  2012 November 	&   3600  & 160 	 &  $4.73\pm0.20$   &	  $5.19\pm0.15$      &  $5.49\pm0.44$ & \\
MMU~3 	    &	2012 November   &   3600  & 150	 &  $4.11\pm0.20$   &   $4.56\pm0.10$     &                            &  \\
MMU~11   	&  2012 November	&   3600  & 175    &  $4.45\pm0.19$	 &   $4.75\pm0.12$      &                           &  \\
MMU~24 	    & 2012 October    &   3600  & 185	 &  $4.86\pm0.19$   &	   $5.36\pm0.10$     &                           &  \\
MMU~27 	    &	2012 November   &   3600 & 190	 &	 $4.39\pm0.19$   &   $4.58\pm0.11$      &                          &  \\
MMU~77 	    &	2012 November   &   3600  & 155	 &	 $4.58\pm0.20$	 &   $5.02\pm0.09$      & 	                        &  $6.3\pm0.2$  \\
MMU~137 	&	2012 October      &   1350  & 170	 &	 $5.59\pm0.20$  &   $5.25\pm0.09$      &                             &  $5.9\pm0.2$\\
MMU~295 	&	 2014 February    &   2700  & 140	 &   $6.32\pm0.23$  &    $5.20\pm0.09$    &                            &  $6.3\pm0.2$\\
MMU~311 	&	2012 October      &   2700  & 180	 &	 $5.19\pm0.19$	  &    $5.79\pm0.09$     &   $6.00\pm0.30$ &  $6.7\pm0.2$\\
MMU~1367 	&	2012 October     &   2700  & 210	 &	 $3.98\pm0.19$	  &    $4.55\pm0.11$     &                          \\
\multicolumn{6}{c}{Non Members}\\
MMU~39 	    &	2012 October      &   3600  & 180  	&	$-21.05\pm0.20$   &  $-21.67\pm0.14$ \\
MMU~215 	&	2012 October      &   1800  & 290	    &	 ~~~$9.51\pm0.24$      &   ~~~$9.29\pm0.13$	\\
\hline
\end{tabular}
\end{minipage}
\end{table*}

We reduced the data by using standard IRAF\footnote{
http://iraf.noao.edu/}
tasks in the \textit{ccdred} package, including 
bias subtraction, flat-fielding, and scattered light subtraction.
Then, the spectra were extracted by using tasks in the \textit{echelle} 
package.
Th-Ar comparison lamp exposures taken at the beginning and end of each 
night were used for wavelength calibration. 
To filter out cosmic-ray and other single-exposure anomalies,
we combined all available integrations for each star.
In combination with the IRAF \textit{telluric} task, we used the 
spectrum of a hot, rapidly rotating star to remove telluric line contamination
(O$_2$ and H$_2$O) from our target spectra.

Our target stars are solar-metallicity RGs and, 
as such, have very line-rich spectra.  
Pure continuum regions 
stretching more than about half an \AA ngstrom
are very difficult to identify even in the yellow-red
spectral regions; this compromises any attempts to make signal-to-noise 
($S/N$) estimates directly from the reduced spectra.  
After performing some numerical tests, we chose to use the photon counts 
of the final co-added one-dimensional spectral orders as primary 
$S/N$ indicators.
We made these calculations at $\lambda$~$\simeq$~6000~\AA, and they are listed
in Table~3.  
The $S/N$ values computed in this manner are in reasonable agreement with 
estimates that we made of point-to-point spectrum flux scatter in 
tiny spectral regions that appear to be free of line absorption.
We also matched the observed spectra with synthetic spectra (see \S\ref{abunds}),
and made $S/N$ estimates 
from the scatter in the \textit{observed minus computed} spectrum differences.
These values also were in reasonable agreement with the primary photon-count
statistics given in Table~3.

We measured the apparent RV shifts of our targets with the 
cross-correlation method provided in the \textit{fxcor} \citep{ftz93} 
routine of IRAF. 
The cross-correlation technique requires a template spectrum. 
In order to avoid error contributions that can arise from rest-frame 
wavelength corrections made to a stellar standard-star spectrum,
we opted against selecting the spectrum of another observed star as
the template.
Instead we created an artificial spectrum (see \S\ref{abunds}) with model
atmospheric parameters similar to those of the NGC~752 programme stars. 
This spectrum was computed for the wavelength range from 5020 to 5990~\AA.  
We then used the task \textit{rvcorrect} given in IRAF's \textit{rv} 
package to transform the geocentric into heliocentric RVs.
We measured the RV for each star after combining individual exposures.  However, we tested the RV scatter on individual exposures for a couple of stars, finding the scatter to be very small, $\sigma \simeq$ 0.05~\kmsec.

Basic observational parameters of our programme 
stars along with measured RVs their associated errors from IRAF's \textit{fxcor} task are listed in Table~3.
In this table we also give the RVs reported in Mer08, Car11 and Red12.
From these data, for NGC~752 we obtain  
$\langle RV\rangle$ =~$4.82\pm0.20$~\kmsec\
($\sigma$~=~0.63~\kmsec), which is in
good agreement with $5.04\pm0.08$~\kmsec\ derived by Mer08.

The RVs of NGC~752 RG members as determined by 
Mer08 and confirmed in this study are not shared by the suggested 
non-members.
For MMU~39, we derived $RV$~=~$-$21.0~\kmsec, about 26~\kmsec\ 
away from the cluster mean, clearly ruling out membership.
For MMU~215, our $RV$~=~9.5~\kmsec\ differs only by 4.7~\kmsec\ 
from the cluster mean, but this still represents a 7$\sigma$ deviation.
Thus stars MMU~39 and MMU~215 do not belong to NGC~752 by this criterion.
Nevertheless, we kept them in our sample to see if their chemical
compositions would provide further information to address the question of cluster 
membership.

\section{Line Lists and Equivalent Widths}\label{lineEW}

An important task in any spectroscopic abundance analysis is 
to create a list of relatively unblended lines that have reliable
transition probabilities.
This is especially important among stars that are cooler than G spectral 
type, because their spectra are composed of overlapping atomic 
and molecular transitions, which adversely affect many potentially 
useful lines.

\subsection{Assembling the atomic and molecular line lists}\label{linelists}

The atmospheric parameter and chemical composition derivations in this work were 
conducted with the current version of the local
thermodynamic equilibrium (LTE) line analysis and synthetic spectrum code MOOG\footnote{http://www.as.utexas.edu/\textasciitilde chris/moog.html} 
\citep{sne73}. Here we discuss the input atomic/molecular line data.
The transitions employed in this study fell into three categories.
\textit{(1)} Unblended Ti and Fe neutral and ionized species 
lines that were used for model atmospheric parameter determinations. 
Abundances from these lines were calculated from their measured equivalent widths ($EW$). 
\textit{(2)} Selected additional unblended atomic species lines that were 
used for relative abundance ratio determinations. 
Again, $EWs$ were used to calculate their abundances.
\textit{(3)} Other atomic and molecular species lines that are heavily 
blended, or have significant hyperfine/isotopic substructure, were
analyzed using spectral synthesis techniques.

Several sources were used to identify unblended lines: 
solar spectrum atlases (\citealt{del73}, \citealt{kur84},
\citealt{wal11}), solar spectrum identifications (\citealt{moo66}), 
the Arcturus spectrum atlas (\citealt{hin05}), and the interactive 
database of high-resolution standard star spectra SpectroWeb
(\citealt{lob11})\footnote{
http://spectra.freeshell.org/spectroweb.html}.
Since it becomes increasingly difficult to determine the continuum 
in bluer spectral regions,
we did not include the lines at wavelengths shorter than 4500 \AA.
We also tried to bypass the regions for which we applied telluric line removal.
As a result, the overall spectral region we used for $EW$
measurements ranges from 4500 \AA \ to 7250 \AA. 
Additionally, we discarded absorption lines with noticeable 
asymmetrical structure, even if we could not find any information about 
line contamination from the solar and Arcturus spectral resources.

As mentioned above, model atmospheric parameters were determined
using transitions of \mbox{Fe\,{\sc i}}, \mbox{Fe\,{\sc ii}}, 
\mbox{Ti\,{\sc i}}, and \mbox{Ti\,{\sc ii}}.
For these species, we imposed additional line strength limits, based
on $EW$ measurements and initial analyses of programme star MMU~77.
Very strong lines in MMU~77, those with $EWs$ higher than 150~m\AA\ 
(reduced widths 
$RW$~$=$ $log~(EW/\lambda)$ $>$~$-$4.6 at $\lambda$~$\simeq$~6000~\AA), 
and very weak lines ($EW$~$<$~10~m\AA\ or $RW$~$<$~$-$5.8), were discarded 
from the list for all stars.
Our final count of lines used for model parameter determinations
includes 62 \mbox{Fe\,{\sc i}}, 12 \mbox{Fe\,{\sc ii}}, 12 \mbox{Ti\,{\sc i}},  
and 5 \mbox{Ti\,{\sc ii}} lines.
These line strength cutoffs were not applied to species with only a few 
available transitions (\eg, \mbox{Na\,{\sc i}} and \mbox{La\,{\sc ii}}).
We used various references for transition probabilities.
The line list for all species, ordered by atomic number and
ionization state, is shown in Table~4 
(full version available online), where we list the wavelengths, 
lower excitation energies, log~$gf$s, and the references for the adopted 
transition probabilities and hyperfine/isotopic splitting. 
Whenever possible, we used a single laboratory-based homogeneous 
transition probability study for a species.
In particular, for \mbox{Ti\,{\sc i}}, \mbox{Ti\,{\sc ii}}, \mbox{Ni\,{\sc i}}, \mbox{La\,{\sc ii}},
\mbox{Nd\,{\sc ii}}, and \mbox{Eu\,{\sc ii}} we used lab data obtained by the
University of Wisconsin atomic physics group.
Frustratingly, there are no recent and comprehensive lab studies for
\mbox{Fe\,{\sc i}} and \mbox{Fe\,{\sc ii}}, the most crucial elements in
any spectroscopic study of stars.
Therefore, we have taken their transition probabilities mostly from 
\citet{obr91}, 
NIST\footnote{http://physics.nist.gov/PhysRefData/ASD/lines\_form.html}
and VALD \citep{kupka00}\footnote{http://vald.inasan.ru/\textasciitilde vald3/php/vald.php} 
(see online version of Table~4). 
For the other species that we analyzed using their measured $EWs$
(\mbox{Si\,{\sc i}}, \mbox{Ca\,{\sc i}}, \mbox{Cr\,{\sc i}}, and 
\mbox{Cr\,{\sc ii}}), transition probabilities were taken from sources noted 
in Table~4. 
Up to 164 lines are potentially available for $EW$ measurements.

\begin{table}
\label{tab4}
 \centering
 \begin{minipage}{75mm}
  \caption{Line list of species.The machine-readable version of the entire table is available in the online journal.}
  \begin{tabular}{@{}lccccc@{}}
  \hline
Species   &  Wave.  &  LEP  &  log \textit{gf}  &  EW / syn   & Ref.   \\
      &   (\AA)  & (eV)   &  &  &   \\
\hline
\mbox{Li\,{\sc i}}	    &	6707.9    &	0	&           $0.17$	&	syn	& Kurucz \\
CH	                       &	4310	   &	 	&	                    	&	syn	 & Mas14\footnote{\cite{mas14}} \\
CH                         	&	4325       &	 	&	                       &	syn	 & Mas14 \\
C$_{2}$	                &	5160       &	 	&  	                   &	syn	 & Bro14\footnote{\cite{brooke14a}}\\
C$_{2}$	               &	5630       &	 	&	                       	&	syn	 & Bro14\\
CN	                       &	8000       &	    &	                     	&	syn	 & Sne14\footnote{\cite{sne14}}\\
\mbox{O\,{\sc i}}      &	6300.31	&	0	        &	$-9.72$ 	&	syn	 & Caf08\footnote{\cite{caff08}}\\
\mbox{Na\,{\sc i}}	&	5682.64	&	2.101	&	$-0.70$	&	syn	 & NIST\\
\mbox{Na\,{\sc i}}	&	6154.23	&	2.101	&	$-1.56$	&	syn	 & NIST\\
\mbox{Na\,{\sc i}}	&	6160.75	&	2.103	&	$-1.26$	&	syn	 & NIST\\
\mbox{Mg\,{\sc i}}	&	5528.41	&	4.343	&	$-0.62$	&	syn	  & Kurucz\\
\mbox{Mg\,{\sc i}}	&	5711.08	&	4.343	&	$-1.83$	&	syn	 & Kurucz\\
\mbox{Mg\,{\sc i}}	&	7811.11	&	5.941	&	$-0.95$	&	syn	 & Kurucz \\
\mbox{Al\,{\sc i}}	    &	6696.02	&	3.14	&	  $-1.35$	   &	syn	 & Kurucz \\
\mbox{Al\,{\sc i}}	 	&	6696.18	&	4.018	&	$-1.58$ 	&	syn	& Kurucz \\
\mbox{Al\,{\sc i}}	 	&	6698.67	&	3.14	&	 $-1.64$  	&	syn	 & Kurucz\\
\mbox{Al\,{\sc i}}	 	&	7835.30	&	4.018	&	 $-0.65$	&	syn	  & Kurucz\\
\mbox{Si\,{\sc i}}	 	&	5488.98	&	5.614	&	$ -1.90$	&	EW	 &Lob11\footnote{\cite{lob11}}\\
\mbox{Si\,{\sc i}}		&	5517.53	&	5.082	&	$-2.61$ 	&	EW	  &VALD\\
\mbox{Si\,{\sc i}}		&	5665.55	&	4.92	&	 $-2.04$	    &	EW	 &NIST\\
\hline
\end{tabular}
\end{minipage}
\end{table}

Fe-group species \mbox{V\,{\sc i}}, \mbox{Co\,{\sc i}} and \mbox{Sc\,{\sc ii}} 
transitions have significant hyperfine substructure, and their transitions 
should not be treated as single lines.
We still derived their abundances from $EW$ measurement, but with the 
blended-line analysis option in the MOOG code.
For our work, we adopted the hyperfine substructure wavelengths and
relative $gf$s from the \cite{kur11}\footnote{
http://kurucz.harvard.edu/linelists.html} 
line compendium.
These species also lack good recent laboratory data. 
Therefore, we determined empirical log $gf$ values for these species 
from a reverse solar analysis. 
We started the reverse analysis of \mbox{Sc\,{\sc ii}} with a transition 
probability taken from \citet{lawler89}, and the initial transition 
probabilities of \mbox{V\,{\sc i}} and \mbox{Co\,{\sc i}} were taken from 
the Kurucz database. 
For this task, we measured $EWs$ from a very high-resolution solar flux 
spectrum (\citealt{kur84}), and forced the total $gf$ values to reproduce
the solar abundances recommended by \cite{asp09} (see \S\ref{initial} for
further discussion of our solar analysis).

The remaining species of interest (C, N, O, \mbox{Li\,{\sc i}}, 
\mbox{Na\,{\sc i}}, \mbox{Mg\,{\sc i}}, \mbox{Cr\,{\sc i}}, \mbox{Sc\,{\sc i}}, \mbox{Cu\,{\sc i}}, 
\mbox{Zn\,{\sc i}}, \mbox{Y\,{\sc ii}}, \mbox{La\,{\sc ii}}, \mbox{Nd\,{\sc ii}}, \mbox{Eu\,{\sc ii}}) 
have complex transitions, caused by their own substructures and/or 
by blending with other absorbers.
For these species, we derived the abundances by spectrum synthesis,
using recent transition probabilities whenever possible.
Special mention should be made here of the availability of new, very 
extensive and accurate laboratory data for molecules that appear in
almost every spectral region of RG stars: 
the $^{12}$C$_{2}$ and $^{12}$C$^{13}$C Swan system \citep{brooke13,ram14}, 
the $^{12}$CH and $^{13}$CH \citep{mas14} G-band, 
$^{12}$CN and $^{13}$CN \citep{brooke14a,sne14}, red and blue systems, 
and MgH \citep{hink13}.
With the new $gf$ line data for these molecules, the accuracy of C and N 
abundance determinations can be significantly improved. 
Furthermore, syntheses of various atomic transitions that are surrounded
by these molecular lines can now be accomplished with greater confidence.

\subsection{\textbf{Equivalent width measurements}}\label{ewmeasure}

We measured $EWs$ in a semi-automated manner, using a 
modified version of the Interactive Data Language (IDL) code that was
introduced by \citet{roe10} and refined by \citet{bru11}. 
The observed line profiles were matched interactively with theoretical 
Gaussian line profiles in most cases, or Voigt profiles 
for some of the stronger lines. 
Central line depths were also recorded for use in estimating 
initial values of effective temperature (\teff) for the programme stars 
via the line depth ratio method discussed below. 
In Table~5 (also available online) we have listed $EWs$ 
measurements of all our target stars.

\begin{table*}
\label{tab5}
 \centering
 \begin{minipage}{120mm}
  \caption{Equivalent width measurements (in m\AA) of the NGC~752 RG members. The machine-readable version of the entire table is available in the online journal.}
  \begin{tabular}{@{}lccccccccccc@{}}
  \hline
  Species  &  Wavelength  &   \multicolumn{10}{c}{MMU}  \\
  &(\AA)& 1 &  3  & 11  & 24  & 27  & 77  & 137  & 295  & 311  & 1367  \\    
\hline
\mbox{Si\,{\sc i}}   &   5488.98   &   32.4   &   31.7   &  33.3   &   34.7   &   35.6   &   32.3   &   34.3   &   34.6   &  35.6   &   35.5      \\
\mbox{Si\,{\sc i}}  &   5517.53   &   22.7   &   21.9   &  22.6   &   22.7   &   24.5   &   23.7   &   26.4   &   29.2   &  28.4   &   29.7     \\
\mbox{Si\,{\sc i}}  &   5665.55   &   61.4   &   59.9   &  62.1   &   26.9   &   64.9   &   64.8   &   66.7   &   61.1   &  67.6   &   67.1      \\
\mbox{Si\,{\sc i}}   &   5701.10   &      &        &       &        &          &        &        &   57.4    &  54.4   &       \\ 
\mbox{Si\,{\sc i}}   &   5793.07   &   58.0   &   52.2   &  59.8   &   59.9   &   61.5   &   58.7   &   62.2   &   62.7   &  63.8   &   61.2   \\
\hline
\end{tabular}
\end{minipage}
\end{table*}

\begin{figure}
  \leavevmode
     \epsfxsize=8cm
      \epsfbox{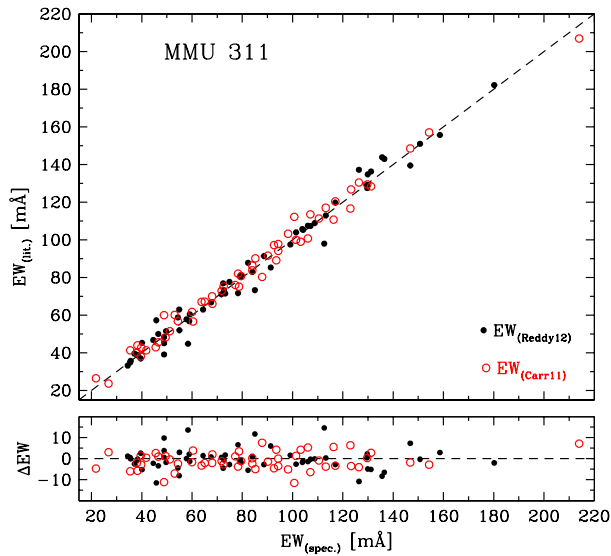}
       \caption{Comparison of our EWs with the ones given by Red12 and Car11 for MMU 311 (top panel).  Differences in EWs are defined as $\Delta$$EW$~$=$~$EW$$_{literature}$~$-~$$EW$$_{this~study}$ (bottom panel).}
     \label{fig2}
\end{figure}

We tested the accuracy of our $EWs$ measurements in several ways.
First, we re-measured the $EWs$ of some lines using the Gaussian fit 
approximation in IRAF's \textit{splot} task; good agreement was found
with our IDL code results.
Second, we compared our $EWs$ with those measured by Car11 and Red12. 
In Figure~\ref{fig2} we show $EW$ correlations for MMU~311, 
the NGC~752 RG  shared  by all three studies. 
For 54 lines used by us and Red12, we find 
$\Delta EW_{Red12}$~= 0.58~$\pm$~0.63~m\AA\ ($\sigma$~=~4.69~m\AA), 
and for 59 lines shared with Car11 we find 
$\Delta EW_{Car11}$~= 0.60~$\pm$~0.60~m\AA\ ($\sigma$~= 4.64~m\AA).
Considering the differences in spectroscopic data and measurement methods 
among these three studies, we regard the $EW$ agreement as satisfactory.

\section{Model Atmospheric Parameters}\label{models}

\subsection{Methods}\label{methods}

We determined parameters for the programme stars with a 
standard approach of calculating abundances from the $EWs$ of Fe and Ti 
neutral and ionized lines, and requiring of these abundances:

\begin{enumerate}
\item for \teff, that there be no difference, on average, in the \mbox{Fe\,{\sc i}}
abundances of low and high excitation ($\chi$) potential lines; 
\item for \vmicro, that there be no difference on average between \mbox{Fe\,{\sc i}}
and \mbox{Ti\,{\sc i}} abundances of weak and strong lines (no apparent trend
with reduced width, $RW$~$=$ $log~(EW/\lambda)$);
\item for \logg, that the mean abundances of neutral and ionized Fe and Ti
lines agree;
\item for [Fe/H], that the value employed in creating the model atmosphere
agrees with the derived value.
\end{enumerate}

These four atmospheric parameters are somewhat coupled, 
\eg, the lowest excitation lines are often the strongest ones.  
Since we have a larger excitation potential range for \mbox{Ti\,{\sc i}}
than the other species, and more \mbox{Fe\,{\sc i}}
than the other species, and more \mbox{Fe\,{\sc i}} and \mbox{Fe\,{\sc ii}} lines than
\mbox{Ti\,{\sc i}} and \mbox{Ti\,{\sc ii}} lines, we gave extra weight to Fe in 
deriving model atmosphere parameters.  
After several trials we adopted a uniform 0.35 weight for Ti lines.
The parameter results proved to be insensitive to the exact weight that
was employed here.

We were mindful of the potential for undesired correlations among the
atmospheric parameters as discussed by \cite{tor12}.
Therefore, we used a semi-automated iterative approach in model derivation,
one that allowed examination of abundance changes caused by each 
alteration in \teff, \vmicro, \logg, and metallicity values.
To accomplish the iterations more efficiently, we used a code that
altered the input model parameters in response to abundance slopes
with $\chi$ and $RW$, and mismatches between neutral and ionized species
or input and output metallicities.
This code is a modified version of one that has been employed in previous
large-sample abundance analyses by \cite{hol11} and \cite{roe14}.
The abundance trends in text and graphical form are available for inspection
in each iteration, so that user judgment can be applied to parameter
changes attempted by the code.
Implementation of this scheme is discussed in \S\ref{finalparams}.

\subsection{Initial parameters}\label{initial}

\begin{table*}
\label{tab6}
 \centering
 \begin{minipage}{140mm}
  \caption{Photometric and spectroscopic atmospheric parameters.}
  \begin{tabular}{@{}lcccccccc@{}}
  \hline
Star   &  T$_{\textit{eff,(B-V)}}$  &  $T_{\textit{eff,(V-K)}}$   & $T_{\textit{eff,(LDR)}}$   &
$\mathrm{log} \ g_{\textit{,phot}}$  &  $T_{\textit{eff,spec}}$   &  $\mathrm{log} \ g_{\textit{,spec}}$  & $\xi_{\textit{spec}}$ &
$[M/H]$    \\
  &  (K) &  (K)  & (K)   &   (cm s$^{-2}$)  & (K)    & (cm s$^{-2}$)  & (km s$^{-1}$)  &     \\                           
\hline
\multicolumn{8}{c}{Cluster Members}    \\
MMU~1       &   4979   &   4888   &   $5038\pm18$   &   2.81   &    5005   &   2.95   &   1.07   &   $-0.07$   \\
MMU~3       &   4891   &   4788   &   $4977\pm21$   &   2.79   &    4886   &   2.76   &   1.10   &   $-0.10$   \\
MMU~11     &   4954   &   4910   &   $5056\pm29$   &   2.74   &    4988   &   2.80   &   1.14   &   $-0.06$    \\
MMU~24     &   4860   &   4810   &   $4991\pm22$   &   2.54   &    4839   &   2.42   &   1.23   &   $-0.09$    \\  
MMU~27     &   4862   &   4798   &   $4991\pm26$   &   2.64   &    4966   &   2.73   &   1.16   &   $-0.04$    \\
MMU~77     &   4824   &   4709   &   $4906\pm21$   &   2.67   &    4874   &   2.80   &   1.15   &   $-0.05$    \\
MMU~137   &   4830   &   4771   &   $4985\pm19$   &   2.60   &    4832   &   2.51   &   1.29   &   $-0.16$     \\
MMU~295   &   4960   &   4892   &   $5062\pm29$   &   2.74   &    5039   &   2.88   &   1.10   &   $-0.05$    \\
MMU~311   &   4810   &   4747   &   $4950\pm20$   &   2.63   &    4874   &   2.68   &   1.24   &   $-0.02$     \\
MMU~1367 &   4844   &   4801   &   $4990\pm32$   &   2.52   &    4831   &   2.42   &   1.22   &   $-0.07$    \\
\multicolumn{8}{c}{Non Members}       \\
MMU~39     &        &       &   $5050\pm39$   &       &   4811   &   2.20    &   1.22   &   $-0.33$     \\
MMU~215   &          &       &   $4671\pm61$   &       &    4350  &   1.81    &   1.29   &   $+0.09$   \\      
\hline
\end{tabular}
\end{minipage}
\end{table*}

\cite{gra91} demonstrated that precise \teff\ values for 
F$-$K dwarf stars could be obtained from calibrations of the central depth 
ratios of absorption line pairs selected to have different responses to 
changes in temperature.
Their basic line depth ratio (LDR) method was expanded in \teff, \logg, 
and metallicity space in subsequent studies (e.g., \citealt{str00}, 
\citealt{gra01}). 
In \cite{afs12} we employed several LDRs to confirm the \teff\ values
derived from \mbox{Fe\,{\sc i}} excitation equilibria for a sample of
red horizontal-branch stars.

A comprehensive examination and re-evaluation of the LDR 
technique was done by \cite{bia07a,bia07b}.
Those authors identified 15 pairs formed from 28 total lines 
in the 6199$-$6274~\AA\ spectral range whose depth ratios are sensitive 
functions of \teff, but relatively insensitive to \logg\ and [Fe/H], at 
least for the metallicities of typical disk stars.
\cite{bia07a} developed cubic polynomials to express \teff\ as 
a function of a depth ratio for each line pair, both for sharp-lined stars
and ones with significant rotational broadening.
As discussed in \S\ref{ewmeasure} we measured as many LDR lines 
(Table~1 of \citeauthor{bia07a}) as possible for each programme star.
We missed only two lines from the \citeauthor{bia07a} list, as they
fell in one of our echelle spectrum inter-order data gaps.
Then we applied the polynomial coefficients for non-rotationally-broadened 
stars (their Table~2) to estimate individual LDR temperatures.
A straight mean of 14 LDRs from 26 lines yielded our final 
\teff (LDR) estimates.
These values are tabulated in Table~6, along with their 
standard deviations and the number of participating LDR pairs.

We also estimated photometric effective temperatures of our programme stars 
from the $B$, $V$, and $K$ magnitudes in Table~2.
For this purpose we used metallicity-dependent \teff\ versus color 
calibrations for giant stars given by \cite{Ram05}.
The temperatures that we calculated for $(B-V)_{0}$ and $(V-K)_{0}$\footnote{
The $K$ magnitude is from 2MASS, and \cite{Ram05} label the color $(V-K_{2})$.}
colors are given in Table~6. 
Metallicity is an important consideration in color-temperature 
calibrations, affecting $(B-V)$ more than $(V-K)$ because the $B$ band suffers
a large amount of line blanketing.
Therefore, since $(V-K)_{\rm0}$ colors are nearly metallicity 
independent, for the initial \teff (phot) we only used the 
computed $(V-K)_{0}$ temperature. 
We then calculated a non-weighted mean of \teff (phot) and \teff (LDR) 
to form the initial \teff\ estimate for the programme stars; 
these also are given in Table~6.

To calculate physical (cluster) gravities we used 
the standard equation,
   
\begin{eqnarray}                                           
\logg_{\star} = 0.4~(M_{\rm V\star} + BC - M_{\rm Bol\sun}) + \logg_{\sun}  \nonumber \\  
+4~{\rm log} (\frac{\teff_{\star}}{\teff_{\sun}})+ {\rm log} (\frac{{\it m}_{\star}}{{\it m}_{\sun}}).
\end{eqnarray}

The adopted solar parameters were $M_{bol}~= 4.75,
\logg~= 4.44$ and \teff~= 5780 K.
Temperature-dependent bolometric corrections were calculated with the 
polynomial formula and coefficients given in Table~1 of \cite{tor10}.
For NGC~752, the absolute magnitudes $M_V$ were computed with the $V$ 
magnitudes of Table~2, and the cluster distance modulus, reddening and turnoff
mass were used as given in Table~1. 

To calculate the physical gravities we varied the stellar masses from 1.5 to 1.95 M$_{\odot}$, 
which is a range from the minimum turnoff mass suggested in the literature to RG mass 
provided by the PARSEC isochrone we applied. 
The results of these experiments always led us to the spectroscopic gravities 
that were expected for the RG stars and they always remained in our uncertainty
limits (see \S\ref{errors}). Therefore, for masses of target stars we adopted $1.5 M_{\odot}$, which is 
the turnoff mass of NGC~752 suggested in \cite{bart07}. 
Previous studies (\S\ref{cluster752}) suggest about solar metallicity for NGC~752, so we used an 
initial value of $[$M/H$] = 0$. Finally, we adopted a microturbulent
velocity that is typical for solar-metallicity RG stars: $\xi$~=1.2~\kmsec.

\subsection{\textbf{Final parameters}}\label{finalparams}

\begin{figure}
  \leavevmode
      \epsfxsize=8cm\epsfbox{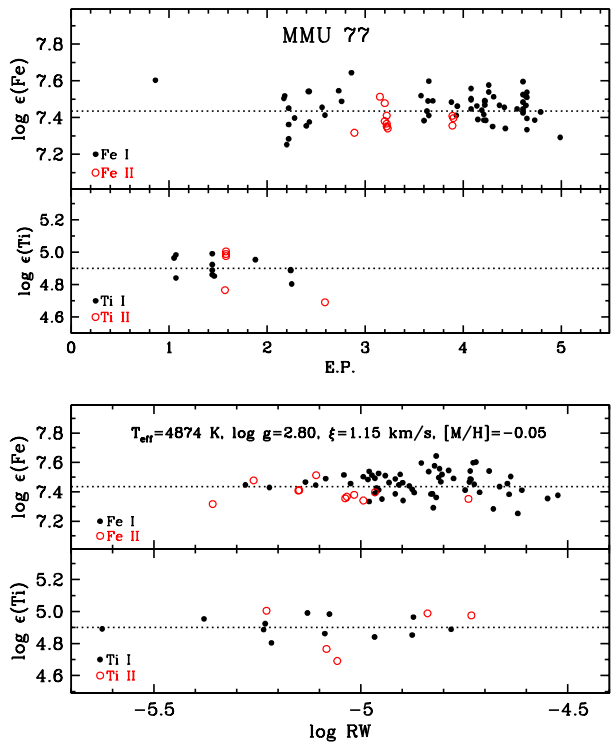}
       \caption{An example model atmosphere determination using both ionized and neutral 
       species of Fe and Ti for MMU 77.}
     \label{fig3}
\end{figure}

We fed the Ti and Fe $EWs$ (\S\ref{ewmeasure}), 
trial model atmospheres, a set of instructions (\eg, model parameter 
goodness of fit criteria), and needed data (\eg, cluster distance
modulus, magnitudes and colors of individual stars) to the semi-automated model 
iteration code.
New grids of ATLAS9 Model Atmospheres from \cite{kur03} were used to compute 
the stellar atmosphere models with opacity distribution 
functions and no convective overshooting. 
These grids were interpolated to the desired
values of \teff, \logg, and [M/H] with software developed by Andy McWilliam 
and Inese Ivans, kindly made available for our use.

After establishing initial model atmosphere parameters, 
we conducted the final parameter search with the code and the model 
convergence criteria described in \S\ref{methods}.
In Figure~\ref{fig3} we plot line-by-line abundance results for star 
MMU~77 after completion of model iterations. 
We will be using MMU~77 in this and several more illustrative plots,
because this star has model atmospheric parameters that are very similar 
to all others in our sample. 
In this figure we have correlated the MMU~77 line abundances with their 
$EP$ and $RW$ values; there is no apparent trend between the abundances and 
these two parameters. 

The initial and final model atmosphere parameters for all of our targets 
are given in Table~6. We also summarize in Table~7 the atmospheric parameters of our programme stars that were obtained by Gil89, Car11, and Red12. 

Comparisons of the final iterated effective temperatures from our spectra, 
$\teff_(spec)$, and the initial \teff\ estimates from photometry and LDR measurements 
are shown in Figure~4. 
On average, values are in reasonable accord, with the mean difference being
$\langle\teff(initial) - \teff(final)\rangle$ = $-$18~K.
However, offsets can be seen among the individual \teff\ determination methods.
For both photometric temperatures, uncertainties arise from reddening and 
distance parameters, and the overall offsets can come from the color-\teff\
calibrations.
We derive $\langle\teff (B-V)-\teff (spec)\rangle = -32\pm13$ and
$\langle \teff (V-K)-\teff (spec)\rangle = -102\pm16$. 
For temperatures calculated with LDR method, 
we find $\langle \teff (LDR) -\teff  (spec) \rangle = 81\pm18$. 
Continuum placement uncertainties can be of concern here, so we
made numerical experiments in which the continuum choices were
changed to substantially lower and higher values.
These resulted in \teff (LDR) changes $\ll$25~K because the LDR method
compares depths of lines situated very close in wavelength, a feature 
central to the original LDR method design \citep{gra91}.
In Figure~4, one can see that the three coolest stars are
offset in the comparisons by $\sim$50$-$70~K compared to the other seven 
stars. This small offset does not substantially effect metallicities and
relative abundance ratios of these stars with respect to the majority
of our sample.

A comparison of calculated gravities (log g$_{phot}$) with spectroscopic 
gravities (log g$_{spec}$) is given in Figure~5.
They agree well with each other:
$\langle \Delta $log g$ \rangle=-0.03\pm0.03$.
There is a small trend that is mostly due to the three coolest, lowest gravity 
RGs in our sample. Differences in the evolutionary status of the members
(see \S\ref{discuss}) may create such a fluctuation.

The metallicities we derived for the NGC~752 RGs from the model atmosphere 
analysis have a slight scatter around the solar metallicity (Table~6). 
The mean metallicity of the cluster calculated from these 10 members is 
$<$[M/H]$>$~= $-0.07\pm0.04$. 
The metallicities of all NGC~752 targets are in agreement except for 
MMU~137, which has a metallicity of [M/H]~= $-0.16$. However, this star
also has one of the lowest \logg\ values in our sample.

We also applied these atmospheric parameter determination methods to the 
suspected non-member stars MMU~39 and MMU~215.
Since no parallax, reddening, and mass data are available for these stars, 
we could not calculate their \teff (B-V), \teff (V-K) and 
log g$_{phot}$ values.
Therefore we used the LDR temperatures for \teff\ and adopted typical RG \logg\
values for the initial parameter guesses.
We found $[M/H]=-0.33$ for MMU~39 ($\sim8 \sigma$ from the cluser mean), 
and $[M/H]=+0.09$ ($\sim3 \sigma$ from the mean) for MMU~215.
Clearly NGC~752 membership is ruled out for these two stars from our
$RV$ and metallicity computations, in agreement with the results of
Mer08. These stars are eliminated from further discussion in this paper.

\begin{figure}
  \leavevmode
      \epsfxsize=8cm
      \epsfbox{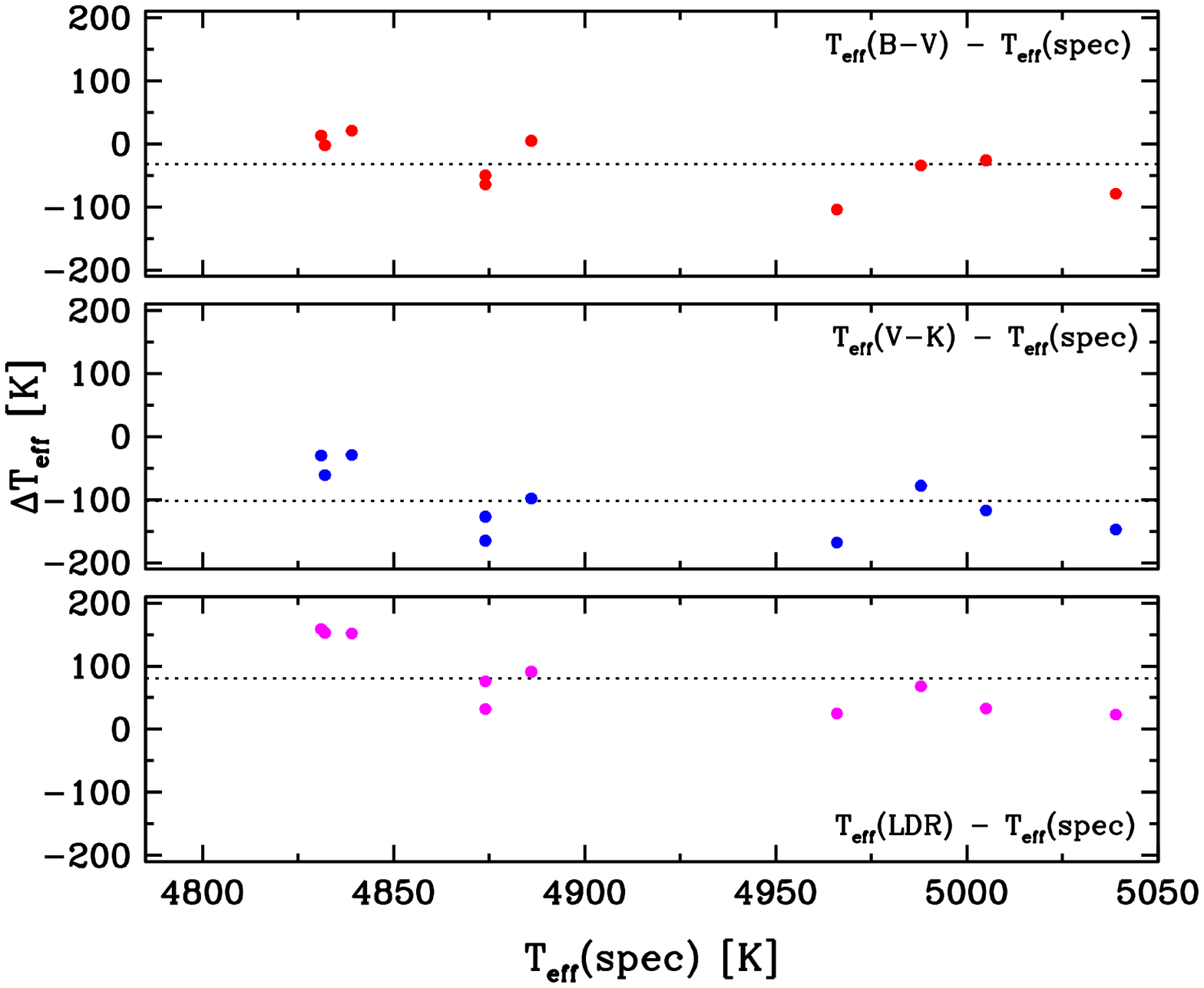}
       \caption{Comparison of $\Delta$\teff$=$\teff($x$)$-$\teff(spec) with spectroscopic $T_{\rm eff}$. $\teff(x)$ stands for the temperatures of $\teff(B-V)$, $\teff(V-K)$ and $\teff(LDR)$.}
     \label{fig4}
\end{figure}

\begin{figure}
  \leavevmode
      \epsfxsize=8cm
      \epsfbox{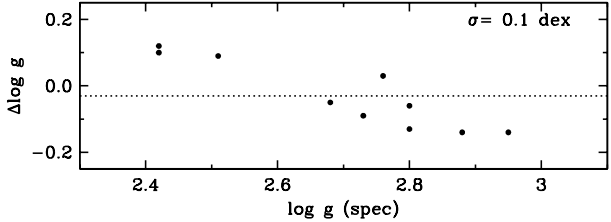}
       \caption{Comparison of $\Delta$log~g$=$log g$_{(phot)}-$log g$_{(spec)}$ with derived spectroscopic log g.}
     \label{fig5}
\end{figure}

\subsection{Parameter uncertainties}\label{errors}

\begin{table}
\label{tab7}
 \centering
% \begin{minipage}{140mm}
  \caption{Spectroscopic atmospheric parameters of NGC~752 members studied by Gil89, Car11 and Red12.}
  \begin{tabular}{@{}lcccc@{}}
  \hline
Star &$T_{\textit{eff,spec}}$   &  $log g_{\textit{,spec}}$  & $\xi_{\textit{spec}}$ &
$[M/H]$      \\
  &  (K) & (cm s$^{-2}$)   & (km s$^{-1}$)      \\                           
\hline
 \multicolumn{5}{c}{Gil89} \\ 
MMU~1       & 5000   &   2.85   &   1.90   &   $0.10$  \\
MMU~77      & 4900   &  2.85    &  1.80    &    $0.20$   \\
MMU~295     & 5000    &  2.90    &  1.80   &   $0.20$     \\
MMU~311     &      4900   &   2.85   &   1.90   &   $0.10$   \\  
 \multicolumn{5}{c}{Car11} \\ 
MMU~1       & 5050   &   3.10   &   1.30   &   $0.045$  \\
MMU~311     &      4800   &   3.20   &   1.20   &   $0.16$  \\  
 \multicolumn{5}{c}{Red12} \\ 
MMU~77        &    4850   &   2.65   &   1.26   &   $-0.03$\\
MMU~137        &  4850   &   2.50   &   1.36   &   $-0.01$\\
MMU~295     &    5050   &   2.85   &   1.47   &   $-0.045$  \\
MMU~311     &    4850   &   2.60   &   1.45   &   $-0.04$ \\  
\hline
\end{tabular}
%\end{minipage}
\end{table}

We estimated the internal uncertainties in atmospheric parameters by running
a series of analyses on the spectral data of MMU~77. First we changed the effective 
temperature in 50 K steps while keeping the other atmospheric parameters fixed.
This process was repeated until the magnitude of the difference between high and low excitation potential
\mbox{Fe\,{\sc i}} lines exceeded the $\sigma$ value of the initial line abundances.   
This method led to an average \teff\ uncertainty of $\sim$100 K.  
We applied a similar method to estimate the internal uncertainty for 
microturbulence velocity. This time we only changed the velocity in 0.1~\kmsec\ 
steps and focused on the abundance changes in the elements that have both 
neutral and ionized lines: \mbox{Ti\,{\sc i}}, \mbox{Ti\,{\sc ii}}, 
\mbox{Fe\,{\sc i}}, \mbox{Fe\,{\sc ii}}, \mbox{Cr\,{\sc i}}.
and \mbox{Cr\,{\sc ii}}.  
The typical average uncertainty achieved for $\xi_{t}$ was 0.3~kms$^{-1}$. 
The internal uncertainty level for \logg\ was also derived by taking into 
account the abundance differences in both neutral and ionized species. 
The highest abundance difference (over $\pm$1$\sigma$ level) between 
\mbox{Cr\,{\sc i}} and \mbox{Cr\,{\sc ii}} was found to be $0.06$~dex, 
which corresponds to an uncertainty of $0.16$~dex in \logg. 

To estimate the external uncertainty in our \teff\ values,
we compared them with the available \teff\ estimates for these stars that were also 
investigated by Gil89, Car11, and Red12.:
Gil89, Car11 and Red12. The atmospheric parameters obtained
in these studies are also listed in Table~7.
We also calculated the photometric 
and LDR temperatures and investigated the differences between these
two temperatures and our spectroscopically derived \teff\ values (Figure~\ref{fig4}). 
An overall comparison of the differences among these temperatures yielded an 
average external uncertainty of about 150 K.

The external uncertainty in \logg\ was estimated by comparing our 
\logg\ values with the ones gathered from the literature for shared NGC~752 stars.
We have made use of the same set of studies that were 
used to estimate the uncertainty in \teff.
Our results are usually in good agreement with the published ones, 
with an average scatter of $\sim$0.2 dex. 
For one star, MMU 311, Car11 reported a \logg\ value of 3.2. 
Since this result is significantly different from the average provided
by other studies, 
we did not include it in our uncertainty estimations.
In Figure~5, we plot the spectroscopic \logg\
versus calculated \logg\ values. 
The standard deviation of the differences between these
two \logg\ values was found to be $\sim$0.1 dex.

Taking into account both internal and external
uncertainty levels, the typical errors we adopted are $\pm$150 K in \teff, 
$\pm$0.25 dex in \logg\ and $\pm$0.3~kms$^{-1}$ in $\xi_{t}$.
We also investigated the uncertainties in elemental abundances caused by these 
adopted uncertainties, and have listed them in Table~8.
The sensitivity of [X/Fe] to the uncertainties in model atmosphere 
parameters is typically much smaller than the values of Table~8
because of correlations in the ionization balances of most species of
interest here.

\begin{table}
 \centering
 \label{tab8}
 \begin{minipage}{90mm}
  \caption{Sensitivity ($\sigma$) of derived abundances to the
   model atmosphere changes within uncertainty limits for the star MMU~77.}
  \begin{tabular}{@{}lccc@{}}
  \hline
Species &  $\Delta$\teff (K) &  $\Delta$\logg  & $\Delta$\vmicro (kms$^{-1}$) \\
 & $-$150 $/$ $+$ 150  &  $-$0.25 $/$ $+$0.25  &  $-$0.3 $/$ $+$0.3   \\
\hline
\mbox{Li\,{\sc i}}	&	$-$0.14	$/$	$+$0.26	&	$+$0.06	$/$	$+$0.09	&	$+$0.09	$/$	$+$0.11	\\
C     	&	$+$0.08	$/$	$+$0.11	&	$+$0.08	$/$	$+$0.09	&	$+$0.09	$/$	$-$0.01	\\
N     	&	$-$0.05	$/$	$+$0.10	&	~~~0.00	$/$	$+$0.05	&	$+$0.03	$/$	$+$0.01	\\
O     	&	$-$0.01	$/$	$+$0.09	&	$-$0.01	$/$	$+$0.19	&	$-$0.01	$/$	$+$0.09	\\
\mbox{Na\,{\sc i}}	&	$-$0.14	$/$	$+$0.07	&	$-$0.01	$/$	$-$0.04	&	$+$0.15	$/$	$-$0.06	\\
\mbox{Mg\,{\sc i}}	&	$-$0.08	$/$	$+$0.05	&	~~0.00	$/$	$-$0.03	&	$+$0.03	$/$	$-$0.07	\\
\mbox{Al\,{\sc i}}	&	$-$0.15	$/$	~~0.00	&	$-$0.10	$/$	$-$0.07	&	~~0.00	$/$	$-$0.10	\\
\mbox{Si\,{\sc i}}	&	$+$0.08	$/$	$-$0.03	&	$-$0.04	$/$	$+$0.05	&	$+$0.04	$/$	$-$0.04	\\
\mbox{Ca\,{\sc i}}	&	$-$0.13	$/$	$+$0.14	&	$+$0.04	$/$	$-$0.04	&	$+$0.15	$/$	$-$0.16	\\
\mbox{V\,{\sc i}}	&	$-$0.24	$/$	$+$0.23	&	~~0.00	$/$	~~0.00	&	$+$0.05	$/$	$-$0.05	\\
\mbox{Cr\,{\sc i}}	&	$-$0.15	$/$	$+$0.16	&	$+$0.02	$/$	~~0.00	&	$+$0.09	$/$	$-$0.09	\\
\mbox{Cr\,{\sc ii}}	&	$+$0.13	$/$	$-$0.08	&	$-$0.10	$/$	$+$0.12	&	$+$0.13	$/$	$-$0.11	\\
\mbox{Sc\,{\sc ii}}	&	$+$0.06	$/$	~~0.00	&	$-$0.08	$/$	$+$0.13	&	$+$0.11	$/$	$-$0.10	\\
\mbox{Ti\,{\sc i}}	&	$-$0.21	$/$	$+$0.19	&	~~0.00	$/$	~~0.00	&	$+$0.08	$/$	$-$0.06	\\
\mbox{Ti\,{\sc ii}}	&	$+$0.07	$/$	~~0.00	&	$-$0.09	$/$	$+$0.13	&	$+$0.12	$/$	$-$0.11	\\
\mbox{Mn\,{\sc i}}	&	$-$0.22	$/$	$+$0.13	&	$-$0.09	$/$	$-$0.02	&	$+$0.05	$/$	$-$0.12	\\
\mbox{Fe\,{\sc i}}	&	$-$0.05	$/$	$+$0.10	&	~~0.00	$/$	$+$0.02	&	$+$0.15	$/$	$-$0.14	\\
\mbox{Fe\,{\sc ii}}	&	$+$0.19	$/$	$-$0.10	&	$-$0.12	$/$	$+$0.15	&	$+$0.09	$/$	$-$0.09	\\
\mbox{Co\,{\sc i}}	&	$-$0.05	$/$	$+$0.12	&	$-$0.02	$/$	$+$0.07	&	$+$0.02	$/$	$-$0.02	\\
\mbox{Ni\,{\sc i}}	&	$-$0.03	$/$	$+$0.10	&	$-$0.03	$/$	$+$0.06	&	$+$0.16	$/$	$-$0.15	\\
\mbox{Cu\,{\sc i}}	&	$-$0.04	$/$	$+$0.13	&	~~0.00	$/$	$+$0.08	&	$+$0.13	$/$	$-$0.09	\\
\mbox{Zn\,{\sc i}}	&	$-$0.04	$/$	$-$0.07	&	$-$0.09	$/$	$+$0.01	&	~~0.00	$/$	$-$0.14	\\
\mbox{Y\,{\sc ii}}	&	$-$0.17	$/$	$-$0.07	&	$-$0.12	$/$	$-$0.07	&	$+$0.06	$/$	$-$0.24	\\
\mbox{La\,{\sc ii}}	&	$-$0.05	$/$	$+$0.01	&	$-$0.11	$/$	$+$0.04	&	~~0.00	$/$	$-$0.04	\\
\mbox{Nd\,{\sc ii}}	&	$-$0.04	$/$	$-$0.02	&	$-$0.15	$/$	$+$0.07	&	$+$0.07	$/$	$-$0.12	\\
\mbox{Eu\,{\sc ii}}	&	$-$0.04	$/$	~~0.00	&	$-$0.09	$/$	$+$0.01	&	$+$0.01	$/$	$-$0.05	\\
$^{12}$C/$^{13}$C	&	2	$/$	0	&	5	$/$	$-$3	&	$-$1	$/$	0	\\
\hline
\end{tabular}
\end{minipage}
\end{table}

We also investigated the uncertainty limits for \ciso~ratios.
Adopting different model atmospheres was not very effective in changing 
the isotopic ratios since the $^{12}$CN and $^{13}$CN molecular lines are 
essentially the same in excitation energies, thus not really sensitive to the changes 
in model atmosphere parameters.
To determine the uncertainties, we fit synthetic spectra with varying \ciso~ratios to 
the observed $^{12}$CN and $^{13}$CN features. 
The resulting uncertainty limits we derived using this method are listed in 
Table~8.

\section{Abundance Analysis}\label{abunds}

Abundances of species with non-blended transitions that could be 
treated as single absorbers were derived from their $EW$ measurements.
Other species exhibiting complex hyperfine splitting 
and/or isotope shifts, and those with lines that suffer significant 
blending of lines by other species, were treated either to blended-line 
analyses or full synthetic-observed spectrum matching. 
Several figures in this section will illustrate the comparisons of 
different synthetic spectra with observed data points.

To normalize our abundances, we re-measured all of
our stellar lines in the integrated solar flux atlas of \cite{kur84}. 
The solar model atmosphere was calculated using the \cite{kur03} grid, 
assuming \teff ~$=$~5777 K, log $g=4.44$ cgs, $\xi=0.85$ km$^{-1}$. 
Abundances found from this analysis are listed in 
Table~9 alongside the solar abundances recommended by \cite{asp09}. 
Since we assumed solar abundances given by \citeauthor{asp09} for         
\mbox{Sc\,{\sc ii}}, \mbox{V\,{\sc i}}, and \mbox{Co\,{\sc i}}, no independent solar 
abundance was derived for these elements.
Estimated sigmas given in parenthesis were determined by taking into 
account the continuum placement, goodness of the fit and smoothing of the 
synthetic spectrum.
For most species, Of course, we employ standard LTE analyses, while more detailed 
physics (e.g., accounting for solar granulation, multi-stream
atmospheric models, non-LTE corrections)
of \citeauthor{asp09} is being neglected.
Even so, our results are mostly in agreement with those of \citeauthor{asp09}.
Our [X/H] and [X/Fe] results for NGC~752 will be quoted differentially
with respect to our own solar analysis abundances.

\begin{table}
 \centering
 \label{tab9}
 \begin{minipage}{60mm}
  \caption{Solar abundances.}
  \begin{tabular}{@{}lcc@{}}
  \hline
Species & log $\epsilon_{\odot}$  & log $\epsilon_{\odot}$     \\ 
 &  (this study)   &  \citep{asp09}       \\                             
\hline
\mbox{Li\,{\sc i}}       &   1.05$\pm(0.05)$			       &   1.05$\pm0.10$     \\
C         &   8.43$\pm(0.05)$  			     &   8.43$\pm0.05$      \\
N         &   8.13$\pm(0.05)$ 			       &   7.83$\pm0.05$       \\
O         &   8.69$\pm(0.05)$       			 &   8.69$\pm0.05$       \\  
\mbox{Na\,{\sc i}}       &   6.34$\pm(0.10)$				        &   6.24$\pm0.04$       \\
\mbox{Mg\,{\sc i}}     &   7.63$\pm 0.16$                      &   7.6$\pm0.04$       \\
\mbox{Al\,{\sc i}}        &   6.33$\pm 0.18$                     &   6.45$\pm0.03$        \\
\mbox{Si\,{\sc i}}        &   7.57$\pm0.05$                       &   7.51$\pm0.03$       \\
\mbox{Ca\,{\sc i}}       &   6.31$\pm0.03$                        &   6.34$\pm0.04$        \\
\mbox{Sc\,{\sc ii}}     &                                     &   3.15$\pm0.04$       \\
\mbox{Ti\,{\sc i}}       &  4.88$\pm0.06$                    &    4.95$\pm0.05$    \\
\mbox{Ti\,{\sc ii}}      &   4.98$\pm0.05$                   &                            \\ 
\mbox{V\,{\sc i}}         &                                      &    3.93$\pm0.08$    \\
\mbox{Cr\,{\sc i}}       &    5.61$\pm0.04$                   &   5.64$\pm0.04$     \\
\mbox{Cr\,{\sc i}}      &   5.72$\pm0.08$                       &                           \\
\mbox{Mn\,{\sc i}}     &   5.41$\pm0.06$                     &    5.43$\pm0.04$    \\
\mbox{Fe\,{\sc i}}    &    7.42$\pm0.04$                   &    7.50$\pm0.04$    \\
\mbox{Fe\,{\sc ii}}     &     7.45$\pm0.04$                     &                             \\
\mbox{Co\,{\sc i}}     &                                       &    4.99$\pm0.07$    \\
\mbox{Ni\,{\sc i}}     &   6.24$\pm0.07$                     &     6.22$\pm0.04$    \\
\mbox{Cu\,{\sc i}}    &    4.07$\pm0.10$                     &    4.19$\pm0.04$    \\
\mbox{Zn\,{\sc i}}      &   4.51$\pm(0.05)$  					      &     4.56$\pm0.05$    \\
\mbox{Y\,{\sc ii}}       &    2.19$\pm0.04$                       &    2.21$\pm0.05$     \\
\mbox{La\,{\sc ii}}      &   1.15$\pm0.06$                       &     1.10$\pm0.04$   \\
\mbox{Nd\,{\sc ii}}      &   1.37$\pm(0.05)$  				       &    1.42$\pm0.04$   \\
\mbox{Eu\,{\sc ii}}      &   0.54$\pm0.08$                    &    0.52$\pm0.04$   \\
\hline
\end{tabular}
\end{minipage}
\end{table}

\begin{table*}
 \centering
 \label{tab10}
 \begin{minipage}{180mm}
  \caption{Elemental abundances of individual stars and average abundances ($<$[X/Fe]$>$) for NGC~752. The average abundances from Red12 and Car11 are also given. The machine-readable version of the entire table is available in the online journal.}
  \begin{tabular}{@{}lrrrrrrrrrrrrr@{}}
  \hline
Species  & \multicolumn{12}{c}{MMU} \\
& 1  & 3  & 11  &  24  &  27  &  77  &  137  &  295  &  311  &  1367  & $<$[X/Fe]$>$ & Red12 & Car11\\ 
\hline
$[\rm{C/Fe}]$	&	$-$0.48	&	$-$0.50	&	$-$0.42	&	$-$0.46	&	$-$0.46	&	$-$0.46	&	$-$0.40	&	$-$0.44	&	$-$0.50	&	$-$0.46	&	$-$0.46	$\pm$	0.03	& ...   &  ...  \\
$[\rm{N/Fe}]$	&	0.15	&	0.07	&	0.10	&	0.14	&	0.18	&	0.12	&	0.14	&	0.14	&	0.13	&	0.11	&	0.13	$\pm$	0.03	&...&...\\
$[\rm{O/Fe}]$  	&	$-$0.21	&	$-$0.23	&	$-$0.21	&	$-$0.20	&	$-$0.13	&	$-$0.13	&	$-$0.15	&	$-$0.15	&	$-$0.18	&	$-$0.20	&	$-$0.18	$\pm$	0.04 & ... &  0.03	\\
$[${\mbox{Na\,{\sc i}}/Fe}$]$	&	0.04	&	$-$0.02	&	0.04	&	0.12	&	0.04	&	0.01	&	0.15	&	0.03	&	0.04	&	0.08	&	0.05	$\pm$	0.05	& 0.12 & 0.01\\
$[${\mbox{Mg\,{\sc i}}/Fe}$]$	&	$-$0.15	&	$-$0.08	&	$-$0.10	&	$-$0.01	&	$-$0.07	&	$-$0.09	&	$-$0.05	&	$-$0.17	&	$-$0.16	& $-$0.09 & $-$0.09 $\pm$ 0.05 & $-$0.01& 0.12	\\
$[${\mbox{Al\,{\sc i}}/Fe}$]$	&	$-$0.02	&	0.02	&	$-$0.02	&	0.07	&	0.03	&	0.05	&	0.10	&	$-$0.01	&	0.00	&	0.04	&	0.03	$\pm$	0.04	& 0.15 & $-$0.12\\
$[${\mbox{Si\,{\sc i}}/Fe}$]$	&	0.06	&	0.08	&	0.08	&	0.15	&	0.09	&	0.13	&	0.19	&	0.09	&	0.10	&	0.15	&	0.11	$\pm$	0.04	&0.11 & 0.02\\
$[${\mbox{Ca\,{\sc i}}/Fe}$]$	&	0.10	&	0.12	&	0.08	&	0.13	&	0.11	&	0.13	&	0.11	&	0.09	&	0.09	&	0.12	&	0.11	$\pm$	0.02	&  0.03 & $-$0.09\\
$[${\mbox{Sc\,{\sc ii}}/Fe}$]$	&	0.00	&	$-$0.03	&	$-$0.01	&	$-$0.02	&	0.01	&	0.01	&	$-$0.01	&	$-$0.02	&	$-$0.01	&	$-$0.06	&	$-$0.01	$\pm$ 0.02	& 0.04 & 0.03\\
$[${\mbox{Ti\,{\sc i}}/Fe}$]$	&	$-$0.03	&	$-$0.06	&	$-$0.08	&	$-$0.08	&	0.04	&	0.03	&	$-$0.04	&	0.00	&	$-$0.04	&	$-$0.09	&	$-$0.03	$\pm$ 0.05 & $-$0.07 &  $-$0.03\footnote{Car11 list the average abundances of the same species, e.g. $[${\mbox{Fe\,{\sc i}}/H}$]$ and $[${\mbox{Fe\,{\sc ii}}/H}$]$ abundances are given as an average: [Fe/H]. Here, we list the abundances from Car11 according to species with the majority of the analyzed lines.} \\
$[${\mbox{Ti\,{\sc ii}}/Fe}$]$	&	0.02	&	$-$0.03	&	$-$0.01	&	$-$0.09	&	$-$0.05	&	$-$0.10	&	$-$0.11	&	0.00	&	$-$0.10	&	$-$0.11	&	$-$0.05	$\pm$ 0.05	& $-$0.04 & ... \\
$[${\mbox{V\,{\sc i}}/Fe}$]$	&	$-$0.09	&	$-$0.11	&	$-$0.15	&	$-$0.14	&	$-$0.04	&	$-$0.02	&	$-$0.10	&	$-$0.09	&	$-$0.10	&	$-$0.17	&	$-$0.10	$\pm$ 0.05	&  0.03 & 0.01 \\
$[${\mbox{Cr\,{\sc i}}/Fe}$]$	&	0.04	&	0.05	&	$-$0.01	&	0.05	&	0.08	&	0.04	&	0.07	&	$-$0.02	&	$-$0.04	&	0.02	&	0.03	$\pm$ 0.04	& $-$0.03 & 0.00 \\
$[${\mbox{Cr\,{\sc ii}}/Fe}$]$	&	0.01	&	0.03	&	0.02	&	0.09	&	0.00	&	0.05	&	0.02	&	0.08	&	0.02	&	0.01	&	0.03	$\pm$ 0.03	& 0.02 &  ...\\
$[${\mbox{Mn\,{\sc i}}/Fe}$]$	&	$-$0.21	&	$-$0.29	&	$-$0.23	&	$-$0.19	&	$-$0.24	&	$-$0.22	&	$-$0.25	&	$-$0.33	&	$-$0.27	&	$-$0.11	&	$-$0.23	$\pm$ 0.06 &$-$0.13 & ...	\\
$[${\mbox{Fe\,{\sc i}}/H}$]$	&	0.04	&	$-$0.05	&	0.03	&	$-$0.05	&	0.08	&	0.04	&	$-$0.08	&	0.07	&	0.07	&	$-$0.02	&	0.01	$\pm$ 0.06	& $-$0.04 & 0.08$^{\textit{\small{a}}}$ \\
$[${\mbox{Fe\,{\sc ii}}/H}$]$	&	$-$0.02	&	$-$0.07	&	$-$0.01	&	$-$0.11	&	$-$0.06	&	$-$0.06	&	$-$0.16	&	$-$0.01	&	0.01	&	$-$0.08	&	$-$0.06	$\pm$ 0.05	& $-$0.02 & ... \\
$[${\mbox{Co\,{\sc i}}/Fe}$]$	&	$-$0.13	&	$-$0.15	&	$-$0.16	&	$-$0.14	&	$-$0.09	&	$-$0.08	&	$-$0.10	&	$-$0.14	&	$-$0.12	&	$-$0.14	&	$-$0.12	$\pm$ 0.03 & $-$0.02 & 0.01\\
$[${\mbox{Ni\,{\sc i}}/Fe}$]$	&	0.01	&	$-$0.03	&	$-$0.01	&	0.02	&	0.07	&	0.06	&	0.03	&	0.05	&	0.03	&	$-$0.01	&	0.02	$\pm$ 0.03	& $-$0.01 &  $-$0.01\\
$[${\mbox{Cu\,{\sc i}}/Fe}$]$	&	$-$0.12	&	$-$0.27	&	$-$0.20	&	$-$0.20	&	$-$0.16	&	$-$0.09	&	$-$0.21	&	$-$0.29	&	$-$0.25	&	$-$0.24	&	$-$0.20 $\pm$ 0.06	& $-$0.11 & ... \\
$[${\mbox{Zn\,{\sc i}}/Fe}$]$	&	$-$0.08	&	0.00	&	$-$0.04	&	0.11	&	$-$0.01	&	0.04	&	0.10	&	$-$0.04	&	0.04	&	0.04	&	0.02	$\pm$ 0.06	& $-$0.10 &  ...\\
$[${\mbox{Y\,{\sc ii}}/Fe}$]$	&	$-$0.09	&	$-$0.09	&	$-$0.01	&	0.01	&	$-$0.07	&	0.02	&	$-$0.13	&	$-$0.08	&	$-$0.11	&	$-$0.13	&	$-$0.06	$\pm$ 0.05	& 0.03 &  $-$0.03$^{\textit{\small{a}}}$\\
$[${\mbox{La\,{\sc ii}}/Fe}$]$	&	0.08	&	0.05	&	0.05	&	0.08	&	0.07	&	0.20	&	0.12	&	0.05	&	0.06	&	0.04	&	0.08	$\pm$ 0.05	&  0.13 &  0.18 \\
$[${\mbox{Nd\,{\sc ii}}/Fe}$]$	&	0.16	&	0.02	&	0.21	&	0.18	&	0.25	&	0.28	&	0.22	&	0.15	&	0.16	&	0.13	&	0.18	$\pm$ 0.07	&  0.06  & 0.34 \\
$[${\mbox{Eu\,{\sc ii}}/Fe}$]$	&	0.01	&	0.13	&	$-$0.05	&	0.10	&	0.01	&	0.12	&	0.09	&	$-$0.02	&	0.05	&	0.03	&	0.05	$\pm$ 0.06	& 0.07 &  ...\\
\hline
\end{tabular}
\end{minipage}
\end{table*}

We derived abundances for 26 species of 23 elements.
We will organize the discussion of these abundances by element groups:
$\alpha$ (Mg, Si, Ca); 
light odd-Z (Na, Al); 
Fe-group (Sc, Ti, V, Cr, Mn, Fe, Co, Ni, Cu, Zn);
$n$-capture (Y, La, Nd, Eu); and 
$p$-capture (Li, C, N, O).
In Table~10 (also available online\footnote{Online version of this table will also include 
the number of lines used for the species and the scatter of the abundance obtained 
from each species.}), we list the derived abundances for individual RG 
members of NGC~752 and the mean abundances for the cluster, $<$[X/Fe]$>$.
The mean abundances from Red12 and Car11 are also given in the last  
two columns of the table for comparison. 
The general distribution of solar normalized elemental abundances 
can be seen in Figure~6. 
The mean abundances of most species (20 out of 26) fluctuate within 
$\pm$0.15~dex of their solar values. 
Abundances of C, Mn, Cu, and Nd depart from the solar abundance mix 
by $\geq$0.2~dex.
We will discuss these and other elemental abundance results in the 
following subsections.

\begin{figure}
  \leavevmode
      \epsfxsize=8.6cm
      \epsfbox{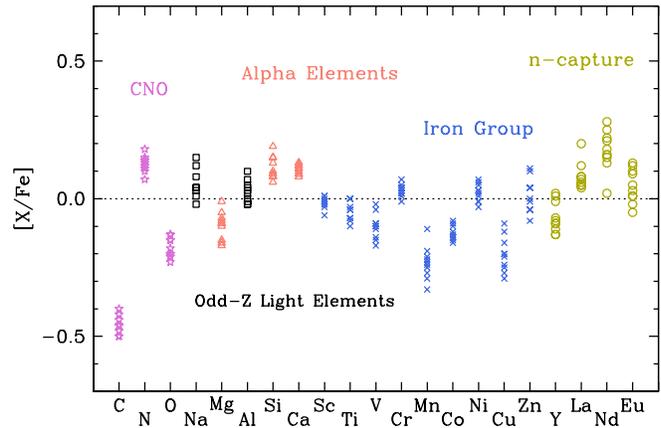}
       \caption{[X/Fe] values of the species studied for the entire RG sample. 
                Dotted line represents the solar values. Abundances of 
                CNO, odd-Z, alpha, iron-group and n-capture elements are 
                shown by stars, squares, triangles, crosses and empty 
                circles, respectively.}
     \label{fig6}
\end{figure}

\subsection{$\alpha$, Odd-Z, Fe-group and Neutron-capture elements\label{alpha}}

\textbf{\textit{$\alpha$ and Odd-Z elements}}: 
The analyzed $\alpha$ elements are Mg, Si, and Ca.
We did not include Ti because even though Ti has been labeled as 
an $\alpha$ element in metal poor stars, 
its dominant isotope $^{48}$Ti$_{22}$ is not an 
even multiple of $\alpha$ particles.
We therefore treat Ti as a Fe-peak.
Si and Ca abundances were derived from $EW$ measurements of their
neutral species lines, while we applied synthetic spectra to determine the Mg 
abundances. Note that three \mbox{Mg\,{\sc i}} lines were used, but the 
feature at 7811.1 \AA\ has very large damping wings that cannot be modeled 
with standard line parameters; the line wings are far too broad.
To compensate, we arbitrarily increased the damping constants for this line 
based on matching it to the solar spectrum.
We also analyzed the 5170 \AA\ region that encompasses two of the very 
saturated \mbox{Mg\,{\sc i}} ``b'' lines, but we did not include their derived 
abundances in the mean Mg abundance. 
We employed spectrum syntheses to determine the abundances of odd-Z 
light elements of interest, Na and Al.

The overall comparison of the average abundances of all species 
with recent NGC~752 high-resolution abundance studies by Red12 and Car11
is given in Table~10. Here and throughout the rest of the paper, our definition of 
comparison is the difference between the mean cluster abundances from these studies and our results:  for Red12, 
$\Delta\rm{[X/Fe]_{R}\equiv[X/Fe]_{Red12}-[X/Fe]_{this \ study}}$, and for 
Car11, $\Delta\rm{[X/Fe]_{C}\equiv[X/Fe]_{Car11}-[X/Fe]_{this \ study}}$.
In Red12, the mean abundances were determined from four cluster members.
Car11 also derived the abundances of four members,
but two of them have been reported as spectroscopic binaries by Mer08.     
Our results are mostly in good agreement with Red12 within error limits:  
$\Delta\rm{[Mg/Fe]_{R}}= 0.06$, $\Delta\rm{[Ca/Fe]_{R}}=-0.08$. 
The discrepancies with Car11 are larger:
$\Delta\rm{[Mg/Fe]_{C}}=0.23$ and $\Delta\rm{[Ca/Fe]_{C}}=-0.20$. 
Our \mbox{Si\,{\sc i}} abundances agree well with those of Red12 
but $\Delta\rm{[Si/Fe]_{C}}=-0.09$.
The differences between the abundances obtained from \mbox{Na\,{\sc i}} 
lines in our study and the others are small, but for the \mbox{Al\,{\sc i}}, 
the differences are $\Delta\rm{[Al/Fe]_{R}}=0.12$ and 
$\Delta\rm{[Al/Fe]_{C}}=-0.15$.
The reason for these discrepancies most likely stem from different analysis 
methods employed in these studies. 
For instance, while we used the spectral synthesis method to obtain
\mbox{Al\,{\sc i}} abundances, Red12 and Car11 used $EWs$.
 
\textbf{\textit{Fe-group elements}}: 
The abundances of \mbox{Ti\,{\sc i}}, \mbox{Ti\,{\sc ii}}, \mbox{Cr\,{\sc i}}, 
\mbox{Cr\,{\sc ii}} and \mbox{Ni\,{\sc i}} were determined from $EWs$ 
in single-line analyses.
As discussed in \S4.1, several Fe-group species (\mbox{Sc\,{\sc ii}}, 
\mbox{V\,{\sc i}}, \mbox{Co\,{\sc i}}) have significant hyperfine 
substructure and no recent comprehensive transition probability studies.
Therefore, we conducted reverse solar analyses to derive 
their transition probabilities. 
Hyperfine component parameters from the Kurucz database were adopted
for these transitions. 
We applied blended-line $EW$ analyses (see \S\ref{linelists}) to 
determine the ``solar'' $gf$'s, and then used these to calculate
the abundances.

We employed full synthetic spectrum computations for other 
Fe-group elements in our list: \mbox{Mn\,{\sc i}}, \mbox{Cu\,{\sc i}}, 
and \mbox{Zn\,{\sc i}}.  
We used three absorption lines of \mbox{Mn\,{\sc i}} located near
$\sim6016$ \AA\ for Mn abundance determination.  
These transitions have hyperfine structure and the region is also 
contaminated by CN absorption. 
We also used three lines of \mbox{Cu\,{\sc i}} to obtain its abundance. 
Although the \mbox{Cu\,{\sc i}} line at 5782.1 \AA\ is
near the spectral gap between the echelle orders, it was taken into account 
whenever it was available. 
The scatter in Zn abundances, determined from only the 
\mbox{Zn\,{\sc i}} line at 6362.3 \AA, is mainly caused by CN and other
atomic contaminants in this region.

Most of our Fe-group abundances are in good agreement with those 
of Red12 and Car11, except for the two reverse solar analysis species, \mbox{V\,{\sc i}} and \mbox{Co\,{\sc i}}.
Although the relative deficiencies that we found in Mn and Cu may be real, 
we repeat our cautions that recent lab studies concerning these elements are unavailable.
For V, the discrepancies are 
$\Delta\rm{[V/Fe]_{R}}\equiv0.13$, 
$\Delta\rm{[V/Fe]_{C}}\equiv0.11$ and for Co, $\Delta\rm{[Co/Fe]_{R}}\equiv0.10$, $\Delta\rm{[Co/Fe]_{C}}\equiv0.13$.
For Mn, $\Delta\rm{[Mn/Fe]_{R}}\equiv0.10$. And for Zn, $\Delta\rm{[Zn/Fe]_{R}}\equiv0.12$. The mean Sc abundance is in good agreement with both Red12 and Car11.

\begin{figure}
  \leavevmode
      \epsfxsize=8cm\
      \epsfbox{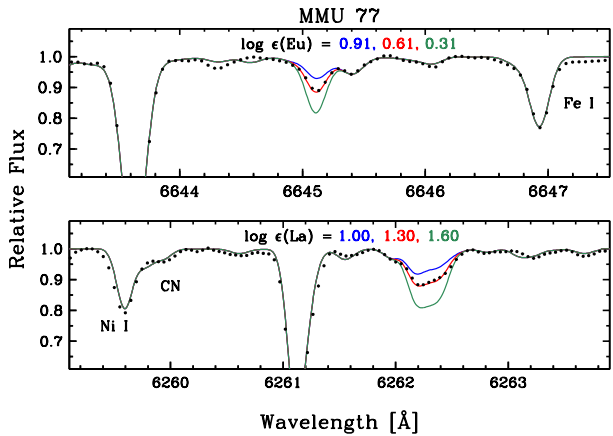}
       \caption{Synthetic and observed spectra comparison of MMU~77 for 
                the 6345.1 \AA\ \mbox{Eu\,{\sc ii}} and 6262.3 \AA\ 
                \mbox{La\,{\sc ii}} lines.
                The observed spectrum is represented by black points.
                Assumed abundances for the three synthetic spectra are
                given in the figure legend.
                The best fit to the observed spectrum is given with a 
                red solid line. 
                The 0.3 dex lower and upper deviations from
                the adopted abundance are represented with green and 
                blue solid lines, respectively.}
     \label{fig7}
\end{figure}

\textbf{\textit{Neutron-capture elements}}: We applied synthetic spectrum 
analyses to three so-called $s$-process (slow neutron-capture)
elements Y, La, Nd, and one $r$-process (rapid neutron-capture) element Eu. 
We also tried to derive Ba abundances in our stars. Unfortunately, all 
\mbox{Ba\,{\sc ii}} lines are saturated in most of our target stars.
Abundances derived from Ba lines are very sensitive to damping, microturbulence,
and the outer atmosphere structures of our stars.
Given these difficulties, we decided to discard Ba from our element list.
Three \mbox{Eu\,{\sc ii}} and four \mbox{La\,{\sc ii}} lines 
were used to determine their abundances. 
In Figure~\ref{fig7}, one of the regions for \mbox{Eu\,{\sc ii}} 
and one for \mbox{La\,{\sc ii}} are illustrated for MMU~77.
We show three different syntheses with assumed abundances that
are separated by 0.3 dex, with the middle (red color) synthetic spectrum
having the abundance that best fits the observed feature.
The synthetic line profiles of these features clearly match the 
details of the observed lines.

All $n$-capture element abundances are in good agreement with those of 
Red12 and Car11 within the mutual uncertainties, except for 
$\Delta\rm{[Nd~II/Fe]_{R}}\equiv-0.12$ and 
$\Delta\rm{[Nd~II/Fe]_{C}}\equiv0.16$. 
The difference may be caused by the 
different analytical methods used in each study.

\begin{figure}
  \leavevmode
      \epsfxsize=8cm
      \epsfbox{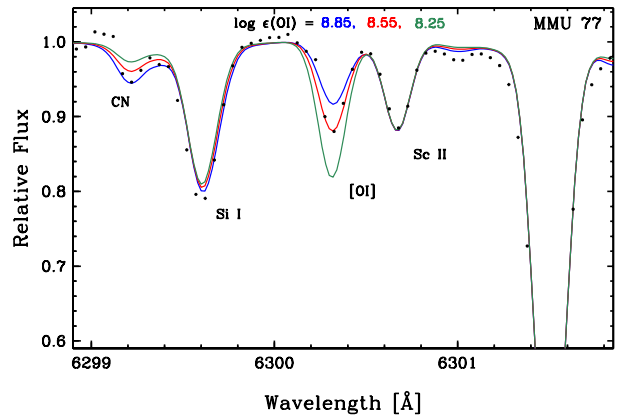}
       \caption{Synthetic and observed spectrum comparison of MMU~77 
                for the [\mbox{O\,{\sc i}}] 6300.3 \AA\ line.
                Lines and symbols are are equivalent to those in Fig. 7.}
     \label{fig8}
\end{figure}

\begin{figure}
  \leavevmode
      \epsfxsize=8.6cm\epsfbox{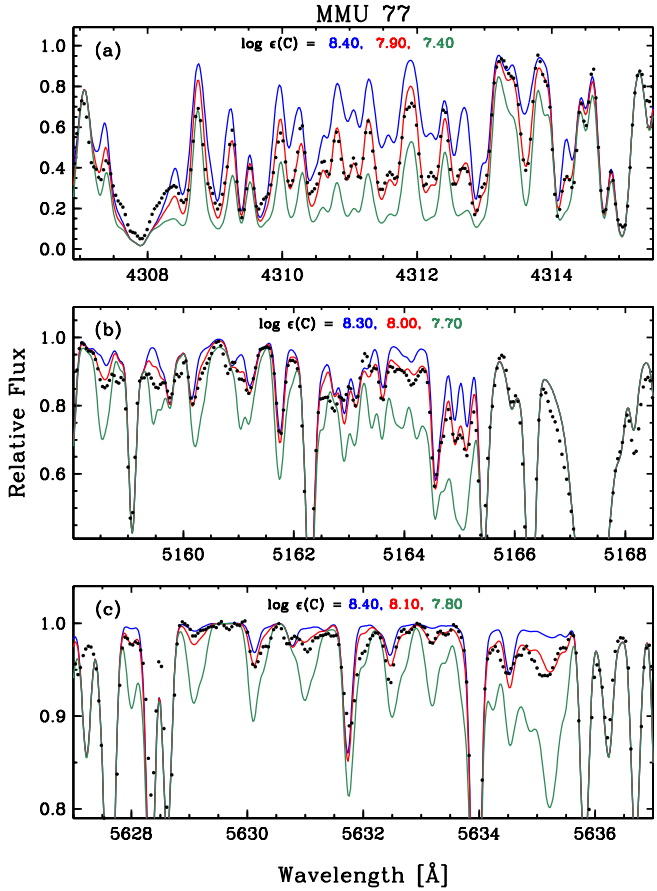}
       \caption{Synthetic and observed spectra of three wavelength regions 
                used for C abundance determination. 
                A portion of the CH G-band (panel a), C$_{2}$ (0$-$0) 
                bandhead at 5155 \AA\ (panel b) and C$_{2}$ (0$-$1) 
                bandhead at 5635 \AA\ (panel c) are illustrated.
                Lines and symbols are as in Figure~\ref{fig7}.}
     \label{fig9}
\end{figure}

\subsection{Proton-capture elements}\label{proton}  

We employed full synthetic spectrum computations to study all features of
the LiCNO abundance group. Using the new molecular laboratory data 
summarized in \S\ref{linelists}, we were able to derive accurate CNO 
abundances for the stars in our sample. 
The elements C, N, and O are interdependent through their coupling into 
molecules such as CN, CH, and especially CO. 
Therefore, their abundances must be determined iteratively, performing 
molecular equilibrium computations at all steps. 
For each star we started our iterations by deriving the O
abundances first from the [\mbox{O\,{\sc i}}] 6300.3 \AA\ line.
Then, C abundances were determined from CH features located around 4311 
\AA\ and 4325 \AA, and C$_{2}$ bands at 5160 \AA\ and 5631 \AA\ region. 
Finally for the N abundances, we used {$^{12}$CN} and $^{13}${CN} 
red system lines in the 7995$-$8040~\AA\ region. 
This process was applied until the changes in element abundances were 
negligible from iteration to iteration.

\begin{figure}
  \leavevmode
      \epsfxsize=8.6cm
      \epsfbox{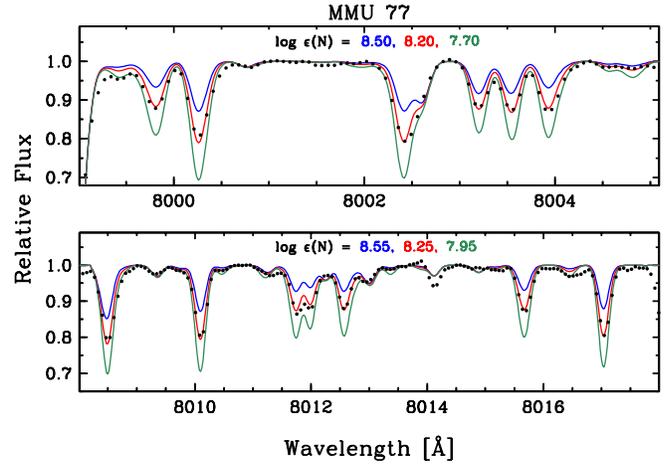}
       \caption{A portion of synthetic and observed spectral regions with CN 
                absorption features used for N abundance determination.
                Lines and symbols are as in Figure~\ref{fig7}.}
     \label{fig10}
\end{figure}

In Figures~\ref{fig8}, \ref{fig9} and \ref{fig10}, we
give examples of synthetic-to-observed spectrum matching for CNO regions 
of MMU~77. 
Figure~8 illustrates the determination of O abundances from 
the [\mbox{O\,{\sc i}}] (6300.30~\AA) line.
The feature is blended with both \mbox{Ni\,{\sc i}} (6300.34~\AA)
and CN (6300.27~\AA).
We tried to account for these contaminants carefully (see \citealt{sne14}
for details). 
Transition probabilities of \cite{joh03} were used for the \mbox{Ni\,{\sc i}} line.
We derived C abundances from several molecular band regions: the CH 
G-band, and the C$_{2}$ (0$-$0) bandhead at 5155 \AA\ and 
the (0$-$1) bandhead at 5635 \AA. 
In Figure~9, we give a portion of CH G-band in panel (a). 
The two bandheads for C$_{2}$ are illustrated in panels (b) and (c).
The complex structures of these molecular regions lead to fluctuating
C abundances from star to star, but they still remained within error limits.
Final C abundances were calculated by taking an average of the values obtained 
from these molecular regions. 
Figure~10 shows an example of N abundance synthesis from the CN region. 
An average was taken to obtain final N abundances.

To our knowledge, the C and N abundances derived here are the first for 
NGC~752 members. 
In standard stellar evolution, the mean N abundance is expected 
to increase and the C abundance decrease by the RGB evolutionary phase, 
Our derived C and N abundances follow this trend. 
However, our O abundances exhibit a subsolar abundance. This is in disagreement 
with the invariance of solar normalized O abundances suggested by classical 
stellar evolution.  
Car11 determined O abundances from the same 
forbidden line and found [O/Fe]=$0.03\pm0.04$. 
However, unlike our study, they did not take into account the interdependence 
of C, N and O abundances while determining the oxygen abundances. 
Furthermore, they found somewhat higher surface gravities (see \S\ref{errors}).
These two effects essentially account for the discrepancy from our result 
of [O/Fe]=$-0.18\pm0.04$. 
Recently, \cite{mad13} reported an average [O/Fe]=$-$0.02 
abundance for the cluster, which was determined from the main-sequence 
members of NGC~752 using 7774 \AA$-${O\,{\sc i}} triplet region. 
Taking into account the non-LTE abundance corrections for oxygen triplet 
(up to almost 0.2 dex, e.g., \citealt{ecu06}), 
Maderak's non-LTE [O/Fe] abundance is found to be around $-$0.22 dex, 
which is in agreement with the oxygen abundance range we report in 
Table~10 and also illustrate in Figure~\ref{fig6}.

\begin{figure}
  \leavevmode
      \epsfxsize=8.6cm\epsfbox{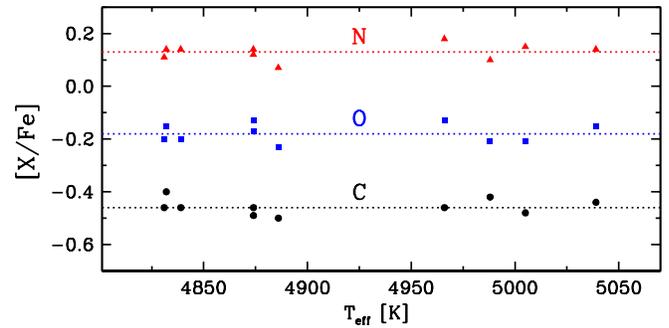}
       \caption{Distribution of C, N and O abundances plotted versus 
                the effective temperatures for RG members.}
     \label{fig11}
\end{figure}

In Figure~\ref{fig11} we plot the derived CNO abundances as 
functions of \teff\ for our 10 NGC~752 programme stars.
This plot reveals no obvious trend with \teff, and there are none to 
be found with the other model atmosphere parameters.

\begin{figure}
  \leavevmode
      \epsfxsize=8.6cm
      \epsfbox{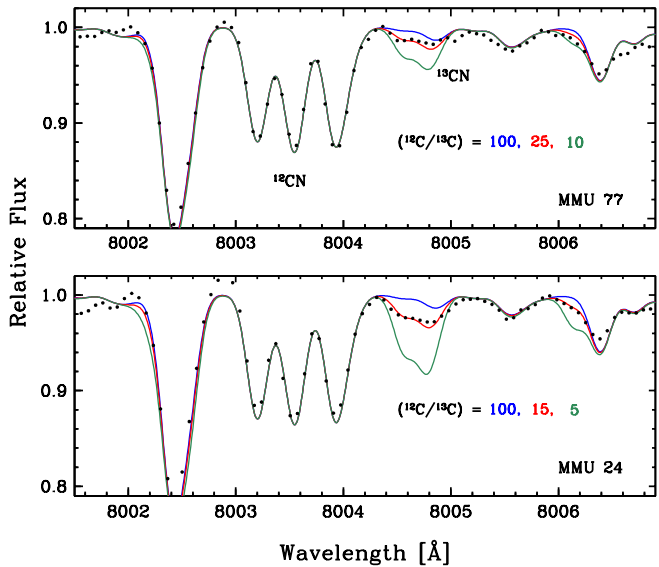}
       \caption{Comparison of synthetic and observed spectra of MMU~77 
                and 24 for the observed CN features around 8004 \AA.  
                These two stars represent the highest and lowest 
                $^{12}$C/$^{13}$C ratios in our sample.
                Lines and symbols are as in Figure~\ref{fig7}.}
     \label{fig12}
\end{figure}

\textbf{\textit{\ciso}}: 
$^{12}$CN and $^{13}$CN features that can be used for 
\ciso\ determination are located throughout the red spectral domain.
The most useful ones for our spectral coverage are in the 
$\sim$8000-8048 \AA\ region. 
Although there are several available features, we decided to determine 
\ciso\ values from the ratio of $^{12}$CN at 8003.5 \AA\ to $^{13}$CN at 
8004.6 \AA\ due to severe line blending in other features. 
In Figure~\ref{fig12}, we 
give an example of different assumed \ciso~ratios matched to this
feature for two members of NGC~752. 
We list all \ciso~ratios for our sample and the ratios gathered from the literature in Table~11. 
The \ciso\ ratios 
determined for our targets range between 13 and 25. 
Gil89 derived \ciso\ for six members of  NGC~752, but later two of them 
were found to be spectroscopic binaries by Mer08. 
The main differences between Gil89 and our work are that we have 
higher spectral resolution and \ciso~ratios were determined during the 
iterative process that we 
obtained CNO abundances, and therefore they should be more reliable. 
Our results indicate that two of the RGs, MMU~1 and MMU~77 have \ciso~=~25 
while Gil89 found lower values:  \ciso=17 for MMU~1 and \ciso=16 for MMU~77.
For the other shared RGs, MMU~295 and MMU~311, there 
is better agreement: we found \ciso=20 and 15, respectively, while 
Gilroy reported \ciso=16 for both members.

\begin{figure}
  \leavevmode
      \epsfxsize=8.6cm
      \epsfbox{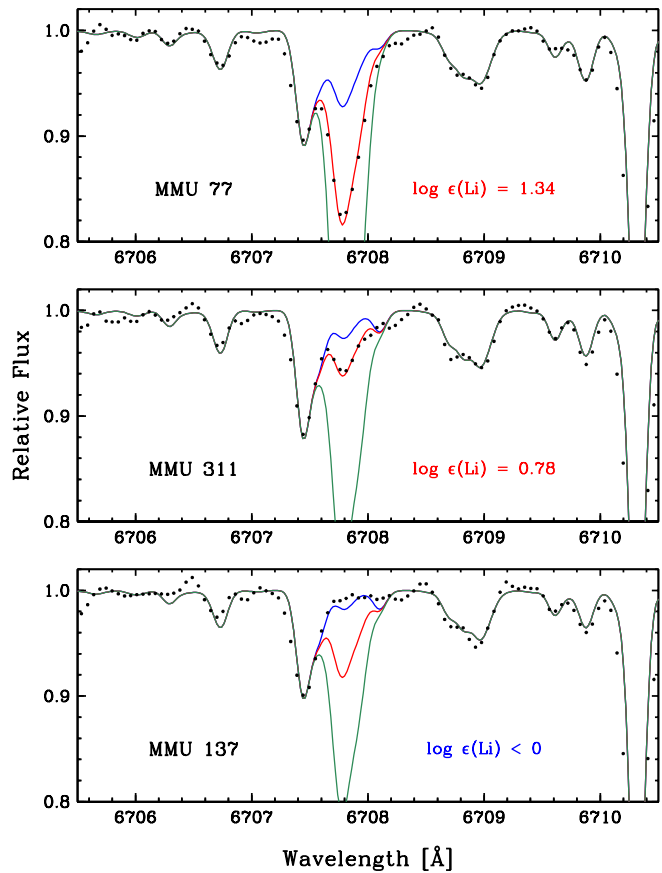}
       \caption{Comparison of synthetic and observed spectra of MMU~77, 
                311 and 137 for 6707.8 \AA\ Li line. These three stars 
                represent the diversity in Li abundances among the RGs 
                of NGC~752.
                Lines and symbols are as in Figure~\ref{fig7}.}
     \label{fig13}
\end{figure}

\textbf{\textit{Li}}: 
We derived Li abundances from the neutral Li 6707.8 resonance doublet, which is somewhat blended with~\mbox{Fe\,{\sc i}} line at 6707.4 \AA. 
Hyper-fine structure of the resonance doublet was also taken into account.
An example of synthetic/observed spectrum comparison is given in 
Figure~\ref{fig13}. 
As seen in this figure, Li abundances vary greatly from star to star.
We were able to measure Li abundances for RGs such as MMU~77 and MMU~311 
(top two panels of Figure~\ref{fig13}), while RGs such as MMU~137 
(bottom panel) does not have a detectable Li feature (log~$\epsilon<$~0).
As tabulated in Table~11, the star-to-star 
abundance range is very large, more than 1.4~dex.
The strongest Li was detected in MMU~77 with an abundance of 
log~$\epsilon$~=~1.34. For the same member, Pil86 and Gil89 also determined Li 
abundances using the same resonance line and found log~$\epsilon$~=~1.1 and 1.4, 
respectively. We could not detect Li in MMU~295, and neither could Pil86 
(log~$\epsilon<+0.5$) but Gil89 reported a detection and with 
log~$\epsilon=0.45$. Our non-detection is to be preferred as it is based on 
higher-quality spectra.

\begin{table*}
 \centering
 \label{tab11}
\begin{minipage}{150mm}
  \caption{Lithium abundances, carbon isotopic ratios and probable evolutionary status of NGC~752 RG members.}
  \begin{tabular}{@{}lcccccc@{}}
  \hline
Star  &  log $\epsilon$(Li)\footnote{\cite{plw88}}   &  log $\epsilon$(Li)\footnote{\cite{gil89}} &   $^{12}C/^{13}C$$^{\textit{\small{b}}}$ & log $\epsilon$(Li)\footnote{This study}        &  $^{12}C/^{13}C$$^{\textit{\small{c}}}$    &  Evol. Phase  \\
\hline
MMU~1 	    & $<$+0.5 &   0.45     & 17        &   0.15 &   25   &     RGB	  \\
MMU~3 	    &         &	           &	       &   1.25 &   25   &     RGB	  \\
MMU~11      &         &            &	       &   1.00 &   25   &     RC	  \\
MMU~24 	    & $<$+0.5 & 	   &	       &   $<$0 &   15   &     RC	  \\
MMU~27 	    &         &	           &	       &   0.95 &   17   &     RC	  \\
MMU~77 	    & 1.1     &	1.40       & 15.5      &   1.34 &   25   &     RGB	  \\
MMU~137     & $<$+0.3 &	           &	       &  $<$0  &   15   &     RC	  \\
MMU~295     & $<$+0.5 &	0.25       & 16        &  $<$0  &   20   &     RHB	  \\
MMU~311     & $<$+0.3 &	0.77       & 16        &   0.78 &   20   &     RC	  \\
MMU~1367    &         &	           &	       &  $<$0  &   17   &     RC	  \\
\hline
\end{tabular}
\end{minipage}
\end{table*}

\section{Discussion and Conclusions}\label{discuss}

This NGC~752 study is the first of a series of papers presenting the 
results of chemical abundance analysis and investigating the 
evolutionary status of the RG members of OCs. 
Our main focus was the abundances of the key elements of 
LiCNO $p$-capture group to help us understand the evolutionary stages of 
the stars and to reveal the discrepancies among RG members.

Before starting with the analysis process, we first confirmed cluster 
membership of the 10 selected RGs in NGC~752 by deriving new $RV$ values.
Since important parameters such as reddening, distance modulus, and turnoff
mass are readily available for clusters such as NGC~752, we were able to 
calculate relatively accurate initial 
parameters for each star.
To derive the most accurate atmospheric parameters, we made a significant effort to 
establish the initial parameters prior to detailed spectrum analysis. 
Initial effective temperatures were determined by taking the average 
of three different temperatures calculated from the $(B-V)$, $(V-K)$ 
colors and line depth ratios of the temperature sensitive species.
We calculated initial log~\textit{g} values using the formula described in 
\S\ref{initial}, and also estimated an initial metallicity of [Fe/H]~=~0. 
Then, we derived final model atmosphere parameters \teff, \logg, \vmicro\ and 
[Fe/H] by using both neutral and ionized species of Fe and Ti.
The model atmospheres of the individual RG members (Table~6)
yielded an average cluster metallicity of about solar, $<$[M/H]$>$=$-$0.07.

We derived the abundances of a number of elements including the light 
(Li, C, N, O), odd-Z (Na, Al), $\alpha$ (Mg, Si, Ca), Fe-group 
(Ti, Cr, Ni, Mn, Cu, Zn) and n-capture elements (Y, La, Nd, Eu). 
We derived \ciso\ ratios of the RGs by employing the synthetic spectrum fitting 
to the {$^{12}$CN} and $^{13}${CN} futures located around 8004~\AA. 
The elemental abundances are mostly scattered around solar values as 
seen in Figure~\ref{fig6}.
From that figure, it is obvious that species such as Cu, Mn, and Nd
exhibit more scatter due to either lack of updated $gf$ values
or complex hyper-fine structure.
However, the overall abundance consistency among all RG members of NGC~752 
indicates that they all share the same origin. 
Recently, \cite{carr14} have reported the abundances of seven RC members
in the OC NGC~4337, which is nearly identical to NGC~752 in age and metallicity.
The CMDs of these clusters are very similar except that NGC~752 has
fewer main sequence stars.  NGC~4337 and NGC~752 appear to have differences in the abundances
of some species.  For example, \citeauthor{carr14} report that Na, V and Co are overabundant
and Zn is underabundant in NGC~4337, while we find that they have almost solar abundances
in NGC~752. CNO abundances in both OCs indicate that the observed members are
certainly evolved. However, C and O abundances are deficient in NGC~752, while N is more
abundant in NGC~4337. These differences suggest that both clusters might have somewhat different
chemical origin. Further study of these species in both clusters is needed to explain and explore these differences
in greater detail.

One of the main goals of our study was to derive reliable abundances of the
$p$-capture LiCNO abundance group.
To our knowledge, our study is the first that reports the C and N 
abundance patterns of evolved stars in NGC~752. 
The results show the agreement among the CNO abundances of the 
individual members, which 
also indicates a common origin for the NGC~752 cluster member RGs. 

The C under-abundances and N over-abundances conform to 
expectations from classical theory of stellar evolution.
$p$-capture reactions are the main mechanisms that govern He-production in 
the core of a star during the main-sequence and later in the H-burning 
shell during the RGB evolution.  
When a star evolves along subgiant branch and becomes a first-ascent RGB
star, its outer convective envelope broadens and reaches down to the 
H-burning shell, eventually dredging the H-processed material up to its
surface.
This so-called first dredge-up phenomenon carries the signatures of internal 
mixing up and shows itself via altered abundances in mainly $^{13}$C and 
$^{14}$N on the stellar surface (e.g. \citealt{iben67, iben84}).

In NGC~752, O is underabundant by about 0.18 dex. 
However, it is not uncommon to find subsolar oxygen abundances 
in OCs. 
We refer the reader to Figure~19 of \cite{yon05}, who
investigate the distribution of [O/Fe] abundances versus age and 
Galactocentric distances (R$_{GC}$) of different OC populations
in the Galaxy. 
NGC~752, with an age of 1.6~Gyr and R$_{GC}$~= 8.3~kpc (Red12),
is similar to many other other OCs with subsolar
metallicities and similar Galactocentric distances.
We also note the small dispersion among the O abundances 
of our cluster members. 
Having the lowest abundance of [O/Fe]=-0.23, MMU~3 has a unique 
spectroscopic feature for our sample: it shows \mbox{Ca\,{\sc ii}} 
HK emission lines. 
MMU~77, which has almost the same temperature and 
gravity as MMU~3, does not carry any signature of \mbox{Ca\,{\sc ii}} 
HK emission, and its [O/Fe]~= $-$0.13.
As suggested in several studies (e.g. \citealt{shc05}), 
stellar atmospheric inhomogeneities such as granulation may be responsible 
for the decrease in the oxygen abundances calculted from the 
forbidden [\mbox{O\,{\sc i}}] line at 6300.3 \AA. 
Further investigation of this 
unusual behavior is beyond the scope of this paper.

According to canonical stellar evolution theories, surface \ciso\ and 
Li content can be altered by the end of the first dredge-up after the main 
sequence and subgiant phases.
They are often the main indicators of stellar evolution. 
We found an unexpectedly large range in both of these abundances 
in NGC~752 RGs.
In our sample, which occupies a relatively small domain of the HR diagram, 
there are examples that have  high \ciso\ and high Li abundances,
indicative of being first-ascent RGBs.
There are also cases with low \ciso\ and Li abundances, indicative of
much further evolution, perhaps the post-He-flash RC evolutionary stage.
We have only one RG, MMU~295, with no Li feature but \ciso~=~20. 
Low \ciso\ ratios have been used as an extra-mixing indicator for some decades.
This explanation dates back at least to Gil89, who found anomalously low 
isotopic ratios in many RG OC members, and showed an anti-correlation
between \ciso\ and the turnoff mass of a cluster.

Our investigation of NGC~752 light elements benefits from higher 
spectral resolution, larger sample size, and new molecular data that yields
CNO abundances along with Li and \ciso.
We found low \ciso\ ratios for only four RGs in in our sample (Table~11). 
Three of these four also show no detection of Li: MMU~24, 137 and 1367. 
Such stars are prime suspects for having undergone some extra-mixing processes. 
However, Li is a fragile element and very sensitive to the details of
main sequence stellar envelope conditions. 
\cite{ses04} have investigated the Li vs. \teff\ distribution in NGC~752
by studying 18 solar-type members and showed that Li abundance decreases as 
the temperature of the cluster members decreases.  This indicates that
the members with slightly different temperatures may have slightly 
different initial Li abundances.

To analyze the light elements among NGC~752 RGs in more 
detail, we decided to explore the evolutionary status of our RG sample.
We re-examined the CMD for NGC~752.
We used the photometric data of Dan94, and applied PARSEC 
isochrones \citep{bress12}, see Figure~\ref{fig1}. 
The best isochrone match to the photometry was found for an
age of 1.60~Gyr and metallicity Z~=~0.014. 
Figure~14 is a zoomed-in version of the CMD Figure~1, centered on the 
RGB and RC regions more in detail. 

We then used the same isochrone to compare our spectroscopically
derived values in the \teff\ - log~$g$ plane for the NGC~752 RGs of this study.
This comparison, which is independent from the photometric parameters,
is displayed in Figure~15.
The locations of the targets in the \teff\ - log~$g$ diagram are in good 
agreement with the ones in Figure~14.

\begin{figure}
  \leavevmode
      \epsfxsize=8.6cm
      \epsfbox{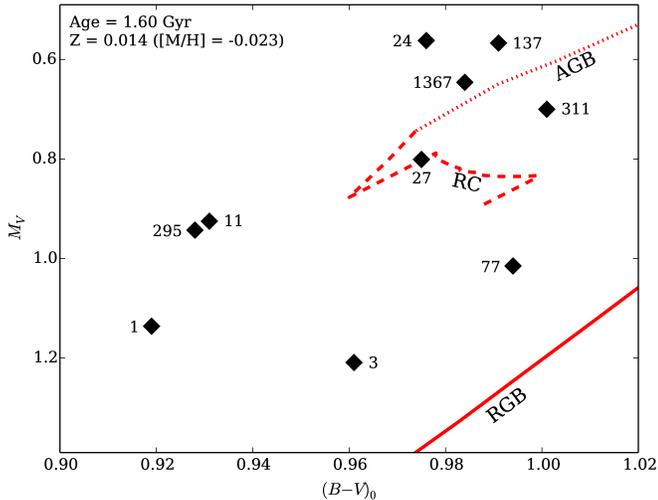}
       \caption{Zoom-in version of the RG region from the CMD given in 
                Figure~\ref{fig1}. 
                The RGB, RC and AGB parts of the 
                isochrone are illustrated with (red) solid, dashed and 
                dotted lines, respectively.
                The points are labeled with the star names.}
     \label{fig14}
\end{figure}

\begin{figure}
  \leavevmode
      \epsfxsize=8.6cm
      \epsfbox{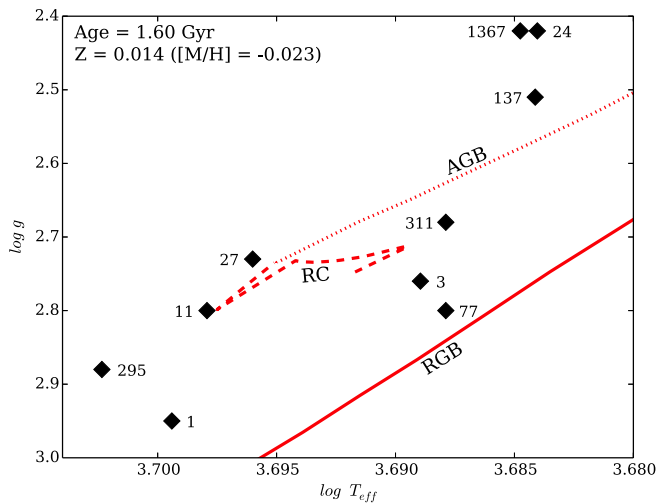}
       \caption{Locations of the RGs in spectroscopic
                \teff - \logg~diagram. Same PARSEC isochrone is 
                used as described in Figure~1. Symbols and labels are the same
                as described in Figure~14.}
     \label{fig15}
\end{figure}

By taking into account the locations of our RGs in both Figures~14 
and 15, along with the Li abundances and \ciso\ ratios derived from the 
spectral analysis, we tried to estimate the probable evolutionary 
stages of our programme stars. 
Our assignments are given in Table~11.
The reader should be warned that we do not indicate \teff\ and
\logg\ uncertainties in Figure~15 for plotting clarity, but they are 
large enough to blur the distinction between RGB and RC membership
in some cases.
Therefore, the evolutionary assignments here are not firm conclusions, 
but rather plausible estimates.

Our reasoning is based on the following observations:
\textit{(1)}: MMU~77 lies closer to the RGB than other RGs in both 
Figure~14 and 15.  
It also has high Li abundance and carbon isotopic ratio (Table~11).
Abundances and CMD location indicate that this star is a regular first-ascent RGB. 
\textit{(2):} MMU~24 and MMU~137 present the most obvious examples of 
post-He-flash RC (or even AGB) stars, as they combine highly evolved 
positions in Figures~14 and 15 with low \ciso\ and no Li detection.
\textit{(3):} The CMD location of MMU~311 appears to indicate RC status.
However, the intermediate abundance values of \ciso\ and Li slightly negate this interpretation.
We label this star as an RC, but caution is advised in this case.
\textit{(4):} MMU~1 is not easily understood.
This star seems to be close to RGB in Figure~15 while it is somewhat more
luminous than a true RGB star in Figure~14. 
It has low Li but high \ciso.
We believe that MMU~1 is an RGB star, but this assignment is not easy to defend.
\textit{(5):} MMU~295 is the hottest RG in our sample. Therefore, it could possibly be 
a red horizontal branch (RHB) star.
The case for this is strengthened by the star's lack of Li detection 
and moderate \ciso~ value of 20. 
We tentatively suggest that our 10-star NGC~752 RG sample contains three 
first-ascent RGBs, six RCs and one RHB.

Spectroscopically-derived accurate temperatures, 
surface gravities, Li abundances and carbon isotopic ratios are 
among the most important parameters to gain better understanding 
about the evolutionary stages of the evolved stars. 
However, matching observed HR Diagram locations of NGC~752 RGs to
isochrones is a challenge without knowing stellar masses.
We have a good estimate of the NGC~752 turnoff mass due to comparisons
of the observed data to both the \logg\ - \teff\ isochrone and the CMD. 
However, individual cluster members might have slightly different initial masses, 
which could lead us to 
observe some diversity among the elemental abundances and \ciso~ratios.
Internal mixing may have taken place with different efficiencies in 
these stars.
As a result, some of the RGs, especially with low \ciso~ratios, may 
have experienced extra-mixing on the RGB after the 1st dredge-up, which takes 
place when low-mass stars reach the so-called luminosity-function bump. 
\citep{grt00,char10}.

Consequently, RGs are important for understanding 
how chemical evolution effects cluster stars. Additionally, they also provide
excellent opportunities to investigate effective mixing 
processes via their \ciso~ratios.
We will present more results that will investigate the chemical evolution of 
the RG members of several OCs in future papers.

\section*{Acknowledgments}

Major parts of this research occurred during several 
exchange visits among the authors at the Department of Astronomy and 
Space Sciences of Ege University and the Department of Astronomy of 
the University of Texas at Austin.
We thank the people of both departments for their hospitality and 
encouragement of this study.
Our work has been supported by The Scientific and Technological 
Research Council of Turkey (T\"{U}B\.{I}TAK, project No. 112T929), by the US
National Science Foundation (NSF, grants AST~09-08978 and AST~12-11585),
and by the University of Texas Rex G. Baker, Jr. Centennial Research Endowment.
This research has made use of NASA's Astrophysics Data System 
Bibliographic Services; the SIMBAD database and the VizieR service, 
both operated at CDS, Strasbourg, France. This research has made use of the WEBDA 
database, operated at the Department of Theoretical Physics and Astrophysics of the 
Masaryk University and of the VALD database, operated at Uppsala University, the 
Institute of Astronomy RAS in Moscow, and the University of Vienna. We would like to 
thank to anonymous referee for his/her helpful discussions and careful review of the 
manuscript.

%\footnotesize{
  %\input{MNRAS.bbl}
 \bibliographystyle{mn2e}{}
 \bibliography{gamze}
%}

\end{document}